\documentclass[screen, authorversion, acmsmall]{acmart}
\usepackage{amsmath}
\usepackage{graphicx}
\usepackage{array}
\usepackage{makecell}
\usepackage{booktabs}
\usepackage[utf8]{inputenc}
\usepackage{natbib}
\usepackage{xcolor}
\usepackage{ifthen}
\usepackage[normalem]{ulem}
\usepackage{svg}
\usepackage{subfiles}
\usepackage{xspace}
\usepackage{cleveref}
\usepackage{wrapfig}
\usepackage{dialogue}
\usepackage[most]{tcolorbox}
\usepackage{hyperref}
\usepackage{hyperxmp}

\ifthenelse{\isundefined{\definition}}{\newtheorem{definition}{Definition}}{}
\ifthenelse{\isundefined{\assumption}}{\newtheorem{assumption}{Assumption}}{}
\ifthenelse{\isundefined{\hypothesis}}{\newtheorem{hypothesis}{Hypothesis}}{}
\ifthenelse{\isundefined{\proposition}}{\newtheorem{proposition}{Proposition}}{}
\ifthenelse{\isundefined{\theorem}}{\newtheorem{theorem}{Theorem}}{}
\ifthenelse{\isundefined{\lemma}}{\newtheorem{lemma}{Lemma}}{}
\ifthenelse{\isundefined{\corollary}}{\newtheorem{corollary}{Corollary}}{}
\ifthenelse{\isundefined{\alg}}{\newtheorem{alg}{Algorithm}}{}
\ifthenelse{\isundefined{\example}}{\newtheorem{example}{Example}}{}

\def\stable{1}

\if\stable1
\else
\fi

\if\stable1
  \newcommand\todo[1]{}

  \newcommand\status[1]{}
  \newcommand\defend[1]{}
\else
  \newcommand\todo[1]{\textcolor{red}{[TODO: #1]}}

  \newcommand\status[1]{\textcolor{gray}{\bf [Status: #1]}}
  \newcommand\defend[1]{\textcolor{magenta}{[Cite: #1]}}
\fi

\if\stable1

\newcommand\pl[1]{}
\newcommand\rb[1]{}
\newcommand\ilevent[1]{}
\newcommand\mx[1]{}
\newcommand\FT[1]{}
\newcommand\tnote[1]{}
\newcommand\dor[1]{}
\newcommand\pw[1]{}
\newcommand\at[1]{}
\newcommand\dora[1]{}

\else

\newcommand\pl[1]{\textcolor{purple}{[Percy: #1]}}
\newcommand\rb[1]{\textcolor{magenta}{[Rishi: #1]}}

\newcommand{\tnote}[1]{{\color{blue}{[TM: #1]}}}

\definecolor{CMpurple}{rgb}{0.6,0.18,0.64}
\definecolor{atcolor}{rgb}{0.83,0.28,0.06}
\definecolor{ballblue}{rgb}{0.13, 0.67, 0.8}

\newcommand\cms{\bgroup\markoverwith{\textcolor{CMpurple}{\rule[.4ex]{2pt}{0.8pt}}}\ULon}

\definecolor{stanford}{rgb}{0.54,0.08,0.08}

\newcommand\mx[1]{\textcolor{orange}{[Michael: #1]}}

\definecolor{mypurple}{rgb}{0.8, 0.18, 1}
\newcommand\dor[1]{\textcolor{mypurple}{[Dor: #1]}}

\newcommand\pw[1]{\textcolor{CMpurple}{[PW: #1]}}
\newcommand\at[1]{\textcolor{atcolor}{[AT: #1]}}

\newcommand\FT[1]{\textcolor{blue}{[Florian: #1]}}

\definecolor{tb12}{rgb}{1.0, 0.13, 0.32}

\newcommand\dora[1]{\textcolor{olive}{[Dora: #1]}}

\definecolor{eamorange}{rgb}{.8,.33,0}

\newcommand\ilevent[1]{\textcolor{orange}{[Isabelle L.: #1]}}

\fi

\usepackage{authblk}

\makeatletter         
\renewcommand\maketitle{
{\raggedright 
\begin{center}
{\Huge \bfseries \sffamily \@title }\\[2ex]
{\@author}\\[2ex] 
\end{center}}}
\makeatother

\renewenvironment{abstract}{%
    \newline
    \itshape
    }
{}

\begin{document}
\title{On the workflow, opportunities and challenges of developing foundation model in geophysics}
\author{\mbox{Hanlin Sheng}\textsuperscript{1}}
\author{\mbox{Xinming Wu}\textsuperscript{1}}
\author{\mbox{Hang Gao}\textsuperscript{1}}
\author{\mbox{Haibin Di}\textsuperscript{2}}
\author{\mbox{Sergey Fomel}\textsuperscript{3}}
\author{\mbox{Jintao Li}\textsuperscript{1}}
\author{\mbox{Xu Si}\textsuperscript{4}}

\affil{\textsuperscript{1}Computational Interpretation Group,  Laboratory of Seismology and Physics of the Earth's Interior, School of Earth and Space Sciences, University of Science and Technology of China\\
\textsuperscript{2}SLB\\
\textsuperscript{3}Bureau of Economic Geology, Jackson School of Geosciences, The University of Texas at Austin\\
\textsuperscript{4}Georgia Institute of Technology
}

\renewcommand{\shortauthors}{Computational Interpretation Group (CIG)}

\maketitle
\noindent 
\vspace{-0.2in}
\begin{abstract}
Foundation models, as a mainstream technology in artificial intelligence, have demonstrated immense potential across various domains in recent years, particularly in handling complex tasks and multimodal data. In the field of geophysics, although the application of foundation models is gradually expanding, there is currently a lack of comprehensive reviews discussing the full workflow of integrating foundation models with geophysical data. To address this gap, this paper presents a complete framework that systematically explores the entire process of developing foundation models in conjunction with geophysical data. From data collection and preprocessing to model architecture selection, pre-training strategies, and model deployment, we provide a detailed analysis of the key techniques and methodologies at each stage. In particular, considering the diversity, complexity, and physical consistency constraints of geophysical data, we discuss targeted solutions to address these challenges. Furthermore, we discuss how to leverage the transfer learning capabilities of foundation models to reduce reliance on labeled data, enhance computational efficiency, and incorporate physical constraints into model training, thereby improving physical consistency and interpretability. Through a comprehensive summary and analysis of the current technological landscape, this paper not only fills the gap in the geophysics domain regarding a full-process review of foundation models but also offers valuable practical guidance for their application in geophysical data analysis, driving innovation and advancement in the field.
\end{abstract}

\clearpage
\tableofcontents
\clearpage

\section{Introduction}\label{sec:introduction}

Geophysics is a scientific discipline that investigates the Earth's interior and its physical properties. By analyzing physical phenomena such as seismic waves, gravitational fields, electromagnetic fields, and geomagnetic fields, geophysicists can reveal the internal structure of the Earth, tectonic activity, and dynamic processes. Moreover, geophysics has significant applications in mineral resource exploration, earthquake forecasting, environmental protection, groundwater detection, and archaeology. 

\begin{figure*}[!htbp]%
\centering
\includegraphics[width=1.0\textwidth]{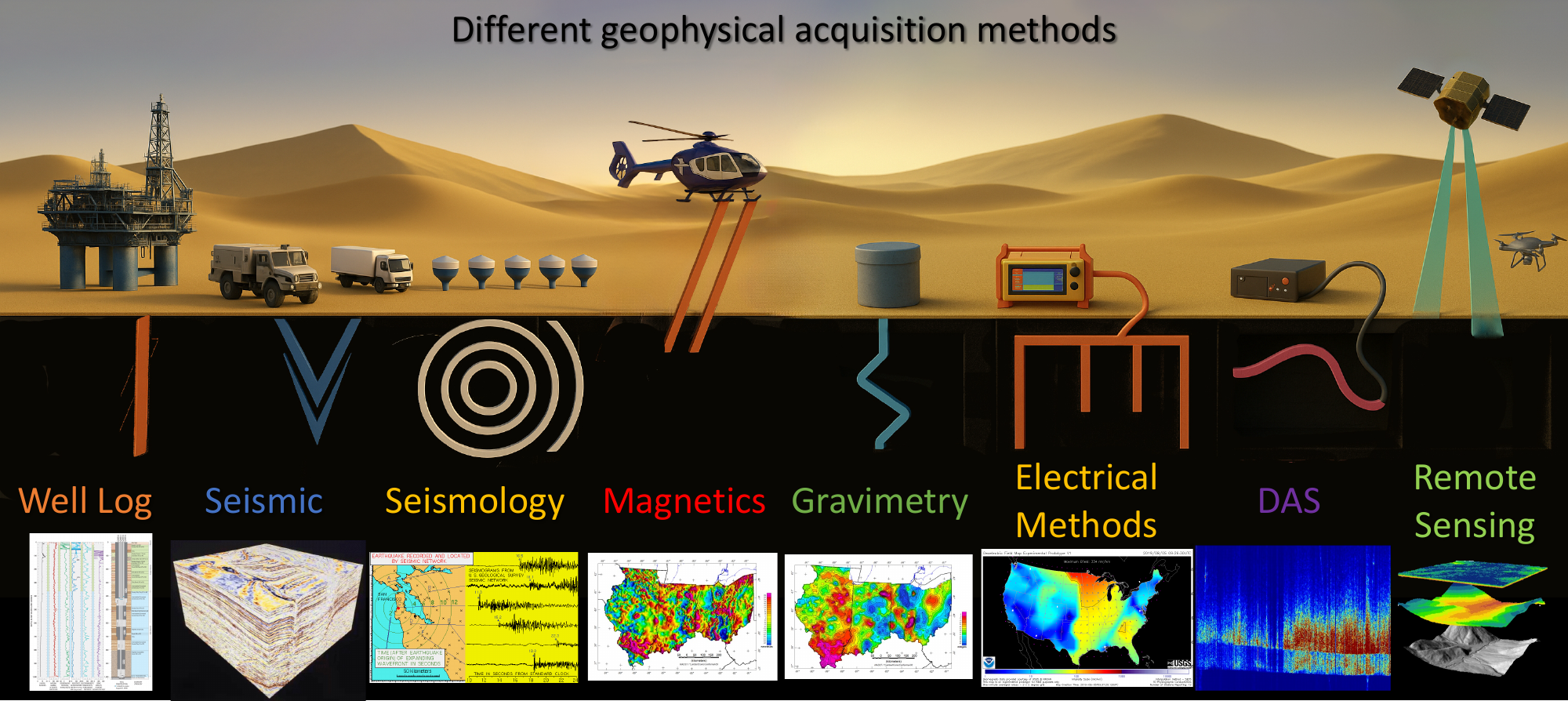}
\caption{Geophysics encompasses a variety of data acquisition methods, including well logging, seismology, magnetics, gravimetry, electrical methods, distributed acoustic sensing (DAS), and remote sensing. The extensive application of these techniques in geophysical exploration, along with their respective advantages, has generated massive volumes of data in diverse formats, spanning various types such as time-series and spatially distributed datasets. With the rapid increase in data volume and the growing diversity of formats, efficiently processing, analyzing, and integrating these datasets has become a critical challenge in contemporary geophysics. (Some images are modified from online sources.)
}
\label{fig:GeoData}
\end{figure*}

There are various geophysical observation and data acquisition methods, each designed to measure different physical quantities to infer information about the Earth's interior (Figure \ref{fig:GeoData}). These methods include drilling, exploration seismology, seismology, magnetics, electrical methods, gravimetry, remote sensing, and distributed acoustic sensing (DAS), among others. Continuous innovation and advancements in observational techniques are key drivers of geophysical progress. Each method has its unique advantages, with different technologies contributing to efficiency improvements and cost reductions, ushering geophysics into a new phase of highly digital development.

With the increasing level of digitization, geophysicists now have access to a vast and diverse array of high-quality data, providing an invaluable resource for scientific research. However, despite the unprecedented quantity and quality of available data, challenges in data processing and analysis remain formidable. Current data processing techniques are often developed for small datasets, making large-scale data processing computationally expensive and time-consuming. The complexity of geophysical data necessitates powerful computational resources and advanced algorithms for effective analysis, posing significant challenges for researchers and engineers. Therefore, improving the efficiency of data processing and analysis while reducing associated costs remains an urgent issue in the field of geophysics.

\begin{figure}[!htb]%
\centering
\includegraphics[width=0.5\textwidth]{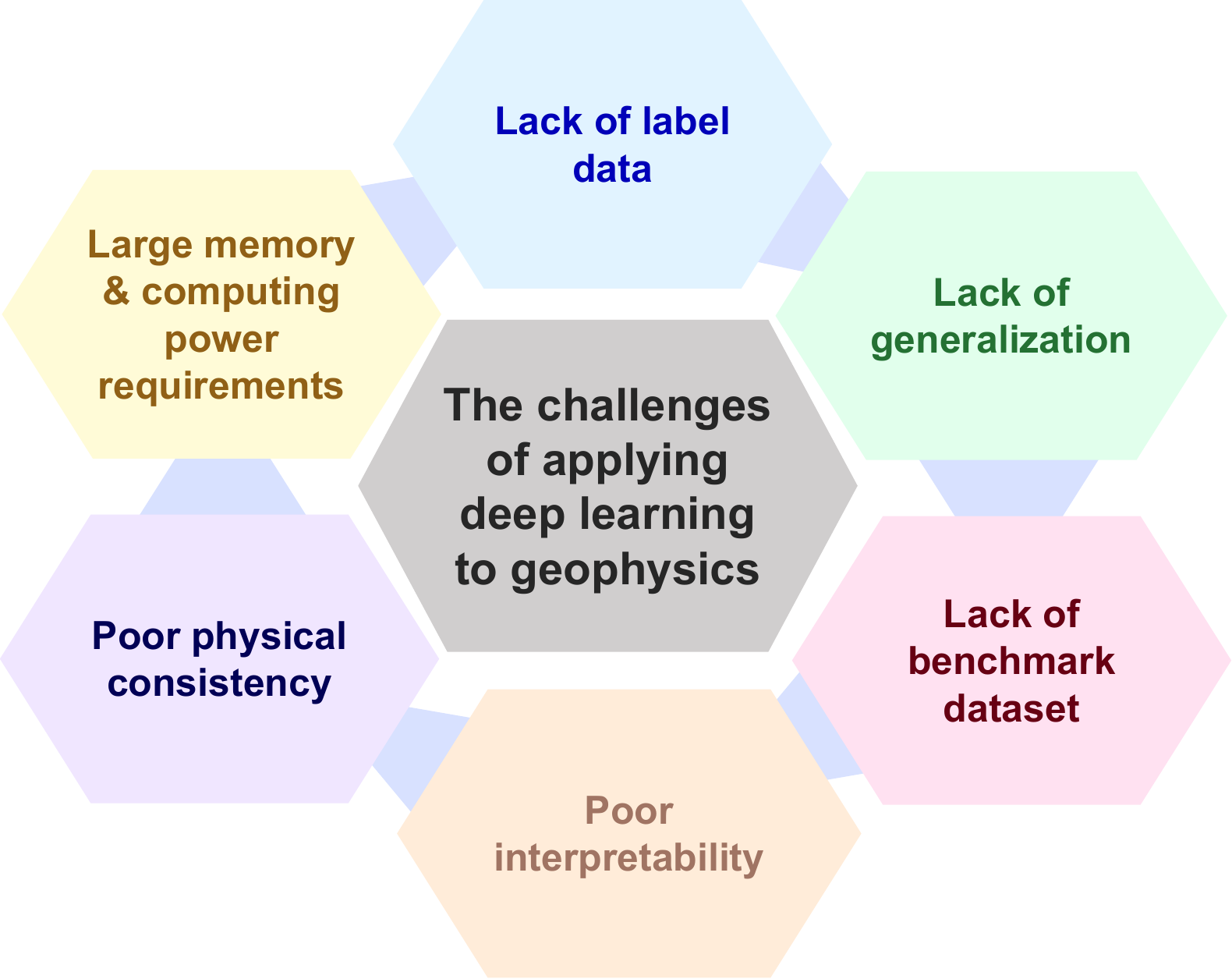}
\caption{The application of deep learning in geophysics faces several major challenges, including the scarcity of labeled data, limited generalization ability, the absence of benchmark datasets, poor physical consistency, low interpretability, and high demands for large-scale memory and computational power. These challenges have constrained the widespread adoption of deep learning in geophysical exploration. Overcoming these limitations requires technological innovations and the development of comprehensive datasets to drive further advancements in this field.
}
\label{fig:challenge}
\end{figure}

In recent years, deep learning has undergone rapid development, particularly driven by the availability of big data and advancements in high-performance computing. Its efficiency and accuracy in data processing and analysis have made it a powerful tool in various scientific domains. Deep learning has demonstrated remarkable success in the geosciences~\cite{bi2023accurate,andersson2021seasonal,lecun2015deep,bergen2019machine,reichstein2019deep,yu2021deep,wu2023sensing}. In geophysics, deep learning has significantly enhanced the analysis of observational data, facilitating the extraction of subsurface rock physics parameters and the investigation of physical processes, thereby deepening our understanding of the Earth's internal structures and dynamics. Specifically, deep learning has played a crucial role across multiple areas of geophysics, including seismology~\cite{mousavi2020earthquake,wang2021seismogen,si2024seisclip,rouet2020probing,yang2022toward,johnson2021laboratory,beroza2021machine,wang2022predicting}, exploration geophysics~\cite{ovcharenko2019deep,park2020automatic,Multiscale2022Feng,harsuko2022storseismic,mousavi2023applications,wu2019faultseg3d,an2023current,geng2020automated,da2021deep, wu2019inversionnet,yang2019deep,chen2020seismic,li2023self,shi2023semi}, well logging applications~\cite{da2021deep,tian2021deep,kanfar2020real}, gravity and electromagnetic exploration~\cite{zhang2021deep,vitale2023deep,liu2020deeper,khan2019deep,jiang2023deep}, geothermal studies~\cite{zuo2019deep,chen2020physics}, and remote sensing~\cite{wang2022comprehensive,zhu2017deep,ma2019deep}. These studies have yielded promising results, demonstrating the effectiveness of deep learning in solving geophysical problems. However, while the application prospects of deep learning in geophysics are broad, several challenges remain in real-world applications, including the scarcity of labeled data, limited generalization ability, the absence of benchmark datasets, poor interpretability, weak physical consistency, and high computational demands (Figure \ref{fig:challenge}). Addressing these challenges is essential to further advancing the integration of deep learning in geophysical research and applications.

One of the primary bottlenecks in the advancement of deep learning in geophysics is the lack of high-quality labeled datasets. This challenge arises mainly due to the difficulty and high cost of data acquisition, as well as the inherent diversity and complexity of geophysical data. Geophysical data are typically collected using sophisticated exploration tools and techniques, such as seismic waves, gravity measurements, and electromagnetic waves. The acquisition of such data often requires operations in extreme or remote environments, making data collection both expensive and challenging.

Moreover, geophysical data exhibit significant heterogeneity due to variations in geological conditions, measurement techniques, and environmental factors, further complicating the annotation process and making standardization difficult. Labeling in geophysics generally relies on expert knowledge, which incurs high costs, and ensuring standardization and consistency in annotations remains a major challenge. Additionally, the lack of interdisciplinary collaboration and unified data-sharing platforms further hinders the construction and integration of large-scale datasets. The slow adoption of advanced labeling techniques, coupled with concerns over data security and privacy, has also limited the development and application of automated annotation methods, exacerbating the shortage of high-quality labeled datasets.

Despite the scarcity of high-quality labeled data, decades of data acquisition efforts have resulted in a vast amount of unlabeled geophysical data, forming a massive repository of raw data. While these data lack precise annotations, they contain abundant latent information that can provide valuable insights into geological, physical, and environmental characteristics. In response to the challenge of acquiring high-quality labeled datasets, researchers are increasingly focusing on methods to extract useful information from these raw data sources. Approaches such as self-supervised learning, unsupervised learning, and innovative data processing and feature extraction techniques are being explored to leverage these vast datasets for constructing more accurate geophysical models. These efforts not only advance the field of geophysics but also create new opportunities for the application of deep learning and big data analytics in geophysical research.

The recent rise of foundation models has brought new hope for addressing key challenges in deep learning, including reducing data requirements, improving computational efficiency, and enhancing model interpretability~\cite{li2023machine}. The concept of foundation models—pretrained on large-scale datasets using self-supervised or semi-supervised learning—has sparked a new revolution in artificial intelligence by enabling adaptability to a wide range of downstream tasks~\cite{bommasani2021opportunities}. These models undergo a single large-scale pretraining process and can subsequently be applied across diverse related tasks.

The motivation behind developing foundation models lies in their superior generalization and transfer capabilities. By leveraging extensive pretraining on vast datasets, these models can efficiently adapt to multiple tasks, thereby reducing the need for task-specific training and significantly lowering computational costs. Foundation models excel in handling complex tasks because they capture richer feature representations and share information across domains, making them particularly advantageous for cross-domain and multimodal applications. Moreover, they effectively utilize large-scale unlabeled data through self-supervised learning, minimizing dependence on labeled datasets. These models have fueled deep learning advancements by promoting the exploration of new methodologies and providing a unified framework for diverse tasks, further driving innovation and efficient data utilization.

The emergence of large-scale foundation models, such as DeepSeek~\cite{guo2025deepseek}, GPT~\cite{openai2023gpt4}, Llama 3~\cite{dubey2024llama}, Gemini~\cite{team2023gemini}, PaLM~\cite{driess2023palm}, Vision Transformer-22B~\cite{dehghani2023scaling}, MAE~\cite{He_2022_CVPR}, CLIP~\cite{radford2021learning}, SAM~\cite{kirillov2023segment}, DINO V2~\cite{oquab2023dinov2}, Sora~\cite{videoworldsimulators2024}, and Whisper~\cite{radford2022robust}, highlights their powerful feature extraction capabilities. These models leverage large datasets, effective training strategies, and well-designed architectures to acquire transferable feature representations, enabling cross-data and cross-task generalization.

While foundation models have achieved significant breakthroughs in computer science, they have also been successfully applied to other fields, including medicine, biology, chemistry, and materials science, where they have facilitated the development of effective solutions. However, research on foundation models in geophysics remains in its early stages. There is a pressing need for in-depth exploration of how foundation models can be integrated with geophysical applications and adapted to the unique challenges of this domain. Addressing these questions will be crucial for advancing the use of foundation models in geophysical research and applications.

Existing studies have conducted some reviews on the application of foundation models in geosciences. For instance, \cite{liu2024research} provides an overview of language models, vision models, and multimodal models in the context of oil and gas exploration geophysics. Similarly, \cite{liu2024foundation} discuss case studies and prospects of foundation models in exploration geophysics, highlighting their applications in seismic data processing. In the field of remote sensing, reviews such as \cite{zhang2024geoscience} explore key technologies related to foundation models and their integration with remote sensing data. However, a systematic review discussing how foundation models can be specifically developed and integrated with geophysical data is still lacking. This gap hinders the widespread adoption and in-depth application of foundation model technology in geophysics.

Current research often focuses on isolated domains and heavily draws from the successes of natural language processing (NLP) and computer vision (CV), but it lacks discussions tailored to the unique characteristics of geophysical data. Critical issues such as preprocessing and modeling strategies for complex, multi-source, multimodal, multi-resolution, and heterogeneous data, the incorporation of physical consistency constraints, and improving the generalization ability of models in geophysical tasks remain underexplored. Therefore, a systematic examination of how foundation models can be adapted and transferred across geophysical disciplines, how model training and applications can be optimized based on data characteristics and physical principles, and how model fine-tuning and multimodal data fusion techniques can be further developed is essential.

This review aims to bridge these gaps by providing a comprehensive summary and analysis to promote the application and development of foundation models in geophysics and its multidisciplinary intersections. The key contributions of this work are summarized as follows:

 \begin{itemize}
        \item \textbf{Systematic framework for constructing geophysical foundation models}\\
This work establishes a complete workflow for developing foundation models in geophysics, covering data collection, preprocessing, model architecture selection, pre-training strategies, and deployment processes. Given the diversity and multi-modal nature of geophysical data, we discuss in detail how to design domain-specific foundation models tailored for geophysics.

		\item \textbf{In-depth exploration of integrating foundation models with geophysical data}\\
Considering the unique characteristics of geophysics, this work analyzes critical strategies for incorporating various geophysical data modalities into foundation model development. In particular, we examine key technologies and methodologies for integrating single-modal and multimodal foundation models, addressing the complexity and heterogeneity of geophysical data.

	\item \textbf{Deployment methodologies and challenges of foundation models in geophysics}\\
This paper provides a detailed analysis of various deployment strategies for foundation models and presents domain-specific optimization recommendations based on geophysical applications. We also compare different model architectures in terms of applicability and generalization performance. Additionally, we discuss deployment costs, open-source culture, funding constraints, and data privacy concerns specific to geophysical data.

	\item \textbf{Future directions and application prospects of foundation models in geophysics}\\
Based on the latest research progress, we outline potential future directions for geophysical foundation models, particularly their potential in interdisciplinary applications and multi-task learning. Through an in-depth analysis of key enabling technologies, this work provides guidance for innovative research on foundation models in geophysics, further advancing the integration of foundation models in this domain.

\end{itemize}

In this review, we focus on the construction and application of foundation models within the field of geophysics. We begin by examining the current state of development of foundation models, providing a summary of general-purpose foundation models, scientific foundation models, and their emerging applications in geophysical contexts. We further analyze the potential of these models in geophysical research, as well as the existing limitations hindering their broader adoption.

Subsequently, we propose a comprehensive framework for building foundation models tailored to geophysical applications, encompassing four key stages: data acquisition, data processing, model training, and model deployment. In the data acquisition stage, we highlight the diversity and strong physical characteristics of both labeled and unlabeled geophysical datasets, and propose strategies to meet the large-scale data demands specific to geophysical modeling.

In the data processing stage, we systematically outline the complete pipeline, including data cleaning, data analysis, data preprocessing, data augmentation, and data evaluation, aiming to ensure data quality that meets the high standards required for foundation model training. For the model training stage, we analyze the advantages of the Transformer architecture, its adaptability to both supervised and unsupervised pretraining, and discuss training strategies tailored to geophysical tasks.

In the model deployment stage, we explore the potential of foundation models for cross-task and cross-regional applications in geophysics, and discuss practical applications and challenges associated with fine-tuning, interactive learning, and other deployment techniques.

Finally, we conduct an in-depth discussion of key issues surrounding the development of geophysical foundation models, including computational resource deployment, open-source collaboration, research investment, and data privacy and security. We conclude by outlining future development directions and challenges, with the aim of promoting further innovation and practical adoption of foundation models in geophysical research.

\section{Current Status of Foundation Model Development}
\label{sec:overview}

\begin{figure*}[!htb]%
\centering
\includegraphics[width=1.0\textwidth]{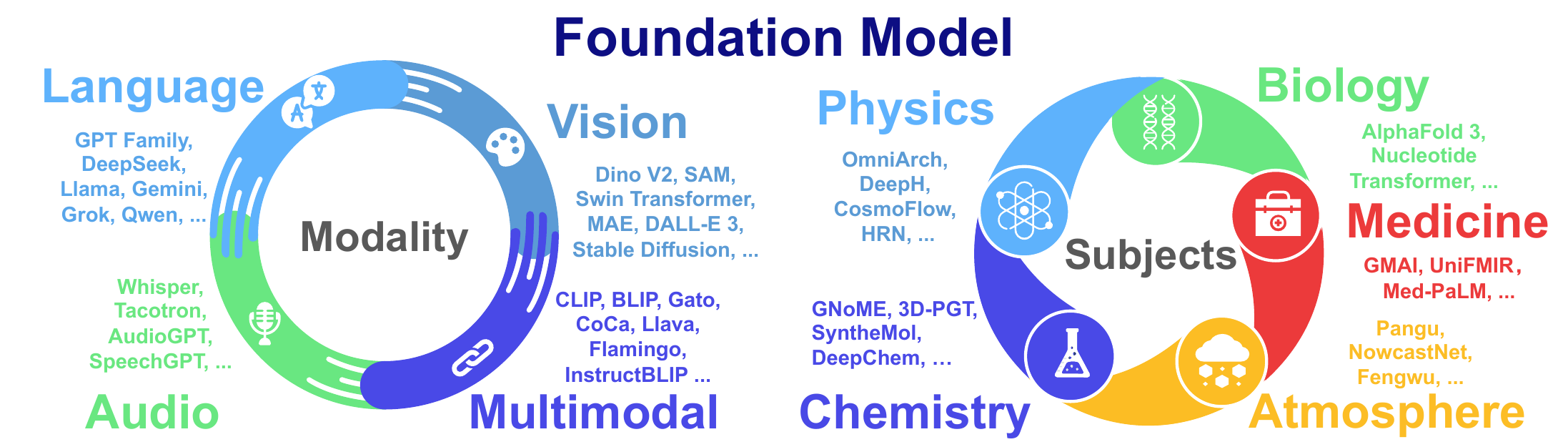}
\caption{
Foundation models have witnessed rapid advancements across language, audio, vision, and multi-modal systems, which in turn have catalyzed progress in various scientific domains such as physics, chemistry, biology, medicine, and atmosphere. This trend underscores the growing influence of foundation models across a wide range of research disciplines.
}
\label{fig:ModelSurvey}
\end{figure*}

In recent years, foundation models have emerged as a transformative breakthrough in the field of artificial intelligence, profoundly reshaping paradigms in both scientific research and engineering practice. By leveraging large-scale pretraining and transfer learning, these models exhibit exceptional generalization capabilities and adaptability to a wide range of tasks. Significant advancements have already been made across disciplines such as medicine, biology, physics, and chemistry (Figure \ref{fig:ModelSurvey}).

At the same time, foundation models have also led to groundbreaking progress in the Earth sciences. In the following sections, we first review the current state of development of general-purpose foundation models, followed by an overview of their applications in scientific domains. We then analyze the distinctive characteristics and challenges faced in the development of foundation models specifically for geophysics, in comparison to other fields. Finally, we provide a summary of recent advances in the application of foundation models within geophysical research.

\subsection{Current Status of General Foundation Model Development}

Recently, artificial intelligence technologies—particularly large language models (LLMs)—have experienced rapid development, driving revolutionary progress in natural language processing, content generation, and code development. The GPT series~\cite{openai2023gpt4}, developed by OpenAI, represents a leading class of proprietary models based on the Transformer architecture, with an estimated parameter count in the hundreds of billions. Renowned for its powerful multimodal capabilities and advanced reasoning performance, GPT has been widely adopted in applications such as intelligent customer service and creative content generation. Llama 3~\cite{dubey2024llama}, released by Meta, is an open-source model available in 8B, 70B, and 405B parameter configurations. It excels in multilingual tasks and code generation, and supports local deployment and customized fine-tuning, offering researchers and developers greater flexibility. DeepSeek-V3~\cite{liu2024deepseek}, developed by the AI company DeepSeek, is another open-source model that employs a Mixture-of-Experts (MoE) architecture with a total of 671B parameters, of which only 37B are activated per inference. It integrates key innovations such as Multi-head Latent Attention (MLA), FP8 mixed-precision training, and Multi-token Prediction (MTP), enabling outstanding performance in Chinese language tasks, mathematical reasoning, and code generation. Claude 3.5 Sonnet, a proprietary model by Anthropic with 175B parameters, demonstrates strong capabilities in creative writing and code generation. Overall, open-source models such as DeepSeek-V3 and Llama 3 are rapidly gaining traction due to their cost-effectiveness and deployment flexibility. Notably, DeepSeek-R1~\cite{guo2025deepseek} has achieved competitive performance on tasks involving mathematics, programming, and logical reasoning, rivaling proprietary models like OpenAI’s GPT-4-turbo (o1). Meanwhile, closed-source models such as GPT and Claude 3.5 continue to lead in complex task performance and multimodal reasoning capabilities.

The foundational models in computer vision and multimodal generative technologies have achieved significant breakthroughs, driving innovative advancements in image and video generation and analysis. Large vision models such as Vision Transformer (ViT\cite{dosovitskiy2020image}), Swin Transformer\cite{liu2021swin}, MAE~\cite{He_2022_CVPR}, and DINO V2~\cite{oquab2023dinov2} have demonstrated remarkable performance in tasks like image classification, object detection, and image generation. Specifically, ViT employs the Transformer architecture with a self-attention mechanism, excelling in capturing global dependencies. Swin Transformer utilizes a hierarchical architecture and shifted window attention mechanism, suitable for large-scale vision tasks. MAE enhances computational efficiency via self-supervised learning and DINO V2 improves feature representation through unsupervised contrastive learning. Additionally, the Segment Anything Model (SAM)\cite{kirillov2023segment} and its upgraded version, SAM2\cite{ravi2024sam}, have provided general-purpose image segmentation solutions without the need for fine-tuning, finding extensive applications in fields such as autonomous driving. Diffusion models, including Stable Diffusion~\cite{esser2024scaling}, DALL-E 3~\cite{ramesh2021zero}, and Imagen Video~\cite{ho2022imagen}, leverage physical diffusion processes to produce high-quality images and videos. Meanwhile, Latent Diffusion Models~\cite{rombach2022high} have enhanced generation efficiency. Furthermore, video generation models like Sora~\cite{videoworldsimulators2024} and EasyAnimate~\cite{xu2024easyanimate} have significantly advanced the development of video generation tasks.

Multimodal large models, such as OpenAI's CLIP~\cite{radford2021learning}, DeepMind's Gato~\cite{reed2022generalist}, BLIP-2~\cite{li2023blip}, and CoCa~\cite{yu2022coca}, integrate multiple modalities including text, images, and audio, significantly advancing cross-modal understanding and generation, and broadening the applications of AI in creative and production domains. For instance, text-to-video generation technologies have automated the creation of advertising videos and movie trailers, simplifying traditional workflows. Specifically, CLIP leverages contrastive learning to embed images and texts within a shared feature space, enabling tasks like image classification and text-image retrieval, showcasing impressive zero-shot learning capabilities and strong generalization. BLIP~\cite{li2022blip} and BLIP-2~\cite{li2023blip} achieve joint representations of images and texts through self-supervised learning and guided pre-training strategies, widely adopted for image captioning and visual question answering tasks. CoCa~\cite{yu2022coca}, by combining contrastive learning and generative modeling, simultaneously handles image-text alignment and text generation tasks, demonstrating excellent performance in multimodal tasks such as image captioning and visual question answering. As these technologies mature, large multimodal models will increasingly automate creative processes and video production, facilitating the intelligent transformation of production tools. This trend is anticipated to bring broader applications across various industries, enhancing efficiency and significantly alleviating human labor.

\subsection{Current Status of Scientific Foundation Model Development}

Pre-trained foundation models have continuously achieved breakthroughs in numerous scientific domains, notably in medicine, life sciences, physics, and chemistry, promoting innovation in analytical methods and research perspectives~\cite{moor2023foundation}. For example, in the medical field, Fudan University's UniFMIR model significantly enhances the restoration quality of biological sample images through pre-training~\cite{ma2024pretraining}, while Alibaba DAMO Academy's PANDA model improves early detection accuracy for pancreatic cancer~\cite{cao2023large}. In biology, AlphaFold3 achieves remarkable accuracy breakthroughs in protein structure prediction~\cite{Abramson2024}, and the Nucleotide Transformer accurately predicts gene phenotypes even in low-data scenarios~\cite{dalla2024nucleotide}. In physics, DeepMind's GNoME model leverages graph neural networks, greatly advancing materials science research~\cite{merchant2023scaling}. In chemistry, the 3D-PGT model pre-trained on three-dimensional molecular structure information performs exceptionally in quantum chemical property prediction and drug screening tasks~\cite{wang2023automated}. Advances in foundational model development not only enhance research capabilities across various disciplines but also provide stronger tools and methodologies for interdisciplinary research. As an important research field, Earth sciences also embrace new development opportunities facilitated by pre-trained foundation models.

In Earth sciences, pre-trained foundation models have achieved significant advancements across several key areas, especially in atmospheric sciences, including weather forecasting, climate simulation, and extreme weather event prediction. For instance, Huawei's Pangu Weather model employs a 3D Transformer architecture trained on large-scale meteorological datasets, significantly improving weather forecasting accuracy and efficiency, and for the first time, comprehensively surpassing traditional physical models~\cite{bi2023accurate}. In predicting extreme weather events, pre-trained foundation models combined with physics-informed deep learning methods notably enhance prediction accuracy and timeliness. For example, Tsinghua University's NowcastNet exhibits outstanding performance in short-term precipitation forecasting, effectively predicting extreme rainfall events~\cite{zhang2023skilful}. The FengWu large model, jointly developed by the Shanghai AI Laboratory and various research institutes, is a global mid-term weather forecasting foundation model employing multimodal and multitask deep learning methods. It provides accurate forecasts extending beyond ten days, and notably requires only a single GPU to produce high-precision global weather forecasts within 30 seconds~\cite{chen2023fengwu}. Given the complexity and diversity of oceanographic data, pre-trained foundation models demonstrate enormous potential in marine environmental monitoring, ocean climate simulation, and marine biological classification. For example, the AI-GOMS model, a representative foundational oceanographic model built on a Fourier operator-based autoencoder architecture, is capable of conducting global forecasts of fundamental ocean parameters such as sea surface temperature, salinity, and current velocity up to 30 days in advance~\cite{xiong2023ai}. Additionally, AI-GOMS extends its downstream applications to biochemical parameters (e.g., chlorophyll concentration, nitrogen content) and wave height prediction, providing novel tools for marine environmental monitoring.

Foundation models have achieved significant advancements across various domains, primarily driven by several critical factors. Firstly, the availability of large-scale datasets constitutes a fundamental driver behind their success. Disciplines such as medicine, life sciences, and physics have accumulated extensive high-quality data sets that not only encompass diverse samples but have also undergone years of meticulous annotation and refinement. Consequently, foundation models can better capture patterns and regularities inherent within the data during training. For example, medical imaging data, genomic data, and meteorological datasets not only possess enormous scale but often feature explicit targets or labels, enabling foundation models to achieve outstanding performance quickly within supervised learning frameworks. Additionally, research in these fields frequently relies upon highly standardized experimental procedures and tools, thereby providing a stable environment for the application of foundation models. Tasks such as medical imaging analysis and protein structure prediction typically involve consistent input formats and target outputs, simplifying dataset construction and enabling efficient, controlled model evaluation processes. However, in interdisciplinary domains, particularly in exploration geophysics, the acquisition and processing of data are confronted with significantly greater complexity and uncertainty.

The field of geophysics holds immense potential, yet its progress in adopting foundation models has been relatively slow due to several critical challenges. Firstly, the scarcity and quality of geophysical data pose substantial obstacles. Data collection related to geology, seismic studies, and subsurface exploration is both difficult and costly, and the resulting datasets often suffer from significant noise and a lack of annotations. Consequently, such datasets are not readily suitable for large-scale model training. Furthermore, geophysical problems typically involve complex physical contexts and strong domain-specific characteristics, requiring traditional physical models to incorporate extensive geological parameters and physical constraints. Therefore, directly applying foundation models in geophysics necessitates close integration with domain-specific knowledge. However, the limited availability of well-labeled data constrains the generalizability and reliability of these models. Additionally, geophysics demands a high degree of accuracy in data annotation. Tasks such as geological exploration and subsurface investigations not only require expensive experimentation and prolonged field observations but also struggle to achieve precise alignment between collected data and specific geological features or physical phenomena, further complicating the development of effective foundation models. Finally, applications within geophysics require model interpretability to ensure alignment with existing physical theories and geological understanding. The inherent ``black-box" nature of deep learning models restricts their applicability in this high-stakes domain. Thus, addressing how to incorporate physical constraints into efficient models, ensuring scientifically credible and trustworthy outcomes, remains an urgent challenge to be resolved.

\begin{figure*}[!htb]%
\centering
\includegraphics[width=1.0\textwidth]{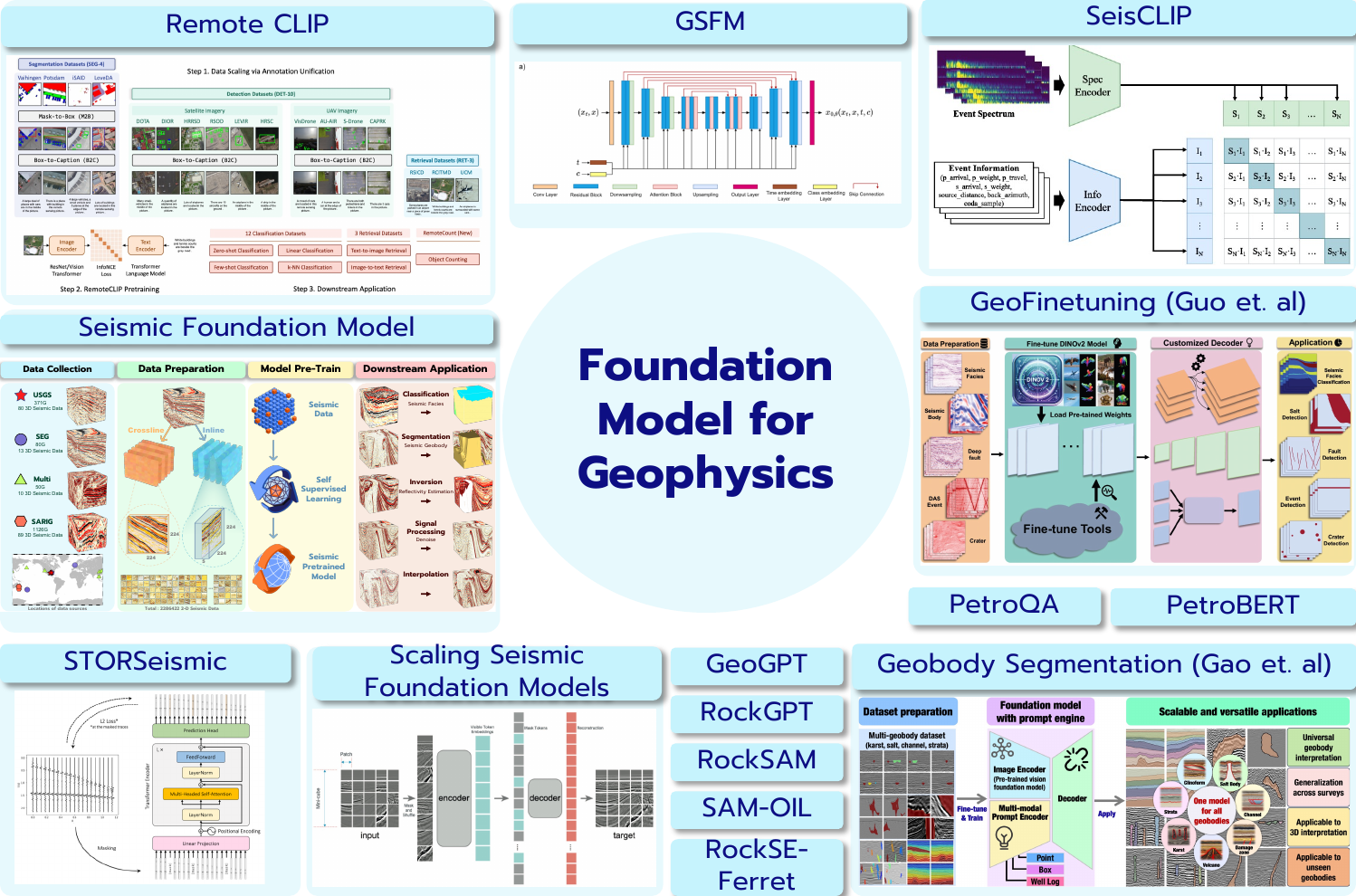}
\caption{In the field of geophysics, several foundation models have recently emerged, including SFM, GSFM, and SeisCLIP. 
}
\label{fig:AI4Sci}
\end{figure*}

\subsection{Current Status of Geophysical Foundation Model Development}

Foundation models have gradually gained deeper adoption in geophysics and remote sensing fields (Figure~\ref{fig:AI4Sci}). \citep{sheng2023seismic} collected over 2.2 million seismic images and introduced a seismic foundation model (SFM) trained using a two-dimensional masked reconstruction pre-training strategy, demonstrating strong generalization and superiority across multiple downstream tasks. \citep{guo2024cross} applied foundational models from computer vision to earth sciences, validating their effectiveness for analyzing diverse geoscientific data, including lunar images, seismic records, and DAS arrays. \citep{gao2024foundation} proposed a tunable prompt-based foundational model for cross-seismic exploration capable of interpreting arbitrary geological bodies. By integrating pre-trained vision foundation models and multimodal prompting engines, this approach achieved high accuracy, scalability from 2D to 3D, strong generalizability, and supported real-time interactivity, marking a new paradigm in geoscientific data interpretation. \citep{chen2024seismic} proposed Seismic Fault SAM, a pioneering attempt to transfer the Segment Anything Model (SAM) into the seismic domain for fault interpretation. By incorporating lightweight Adapter modules, freezing most pretrained weights, and adopting a 2.5D input strategy, this work enables efficient convergence and spatial pattern learning while maintaining computational efficiency.  \citep{ atolagbe2025towards} introduced FaciesSAM, a prompt-enabled seismic facies segmentation framework that leverages the Fast Segment Anything Model (FastSAM) to address longstanding challenges in interpretability, interactivity, and data efficiency. By formulating the task as a Segment All or Segment One (SASO) problem, this approach enables both broad-scale all-instance segmentation and fine-grained prompt-guided facies identification, offering seismic interpreters flexible control in localized analysis. This work highlights the practical advantages of prompt-based segmentation paradigms in geoscientific applications, especially under limited data scenarios.

In seismic signal processing, \citep{harsuko2022storseismic} leveraged BERT-inspired methods for recovering randomly missing seismic traces, enabling the model to effectively extract pre-stack data features. After further pre-training with real seismic datasets, their method demonstrated superior performance across multiple downstream tasks. \citep{liu2024foundation} applied the Segment Anything Model (SAM) to significantly enhance tasks such as first arrival picking, and discussed the integration of language models to develop GeoFM. Additionally, \citep{sansal2025scaling} employed self-supervised pre-training strategies on 3D seismic data to create generalizable foundation models specifically tailored for seismic interpretation tasks. Furthermore, \citep{cheng2025agenerative} proposed the Generative Seismic Foundation Model (GSFM), utilizing generative diffusion models (GDMs) to address common challenges in seismic data processing, including noise contamination, incomplete data acquisition, and insufficient low-frequency information. By conducting pre-training on synthetic data and employing iterative fine-tuning strategies, GSFM achieved outstanding performance in denoising, backscatter noise attenuation, interpolation, and low-frequency extrapolation tasks. Notably, GSFM can effectively quantify uncertainties in results, enhancing performance and versatility in multi-task seismic data processing.

In the domain of seismology, the SeisCLIP model represents a notable foundation model designed specifically for seismic applications. Pre-trained through contrastive learning across multimodal seismic data, SeisCLIP can be effectively fine-tuned on small datasets for diverse downstream tasks, including seismic event classification, earthquake localization, and focal mechanism analysis~\cite{si2024seisclip}. In 2024, \citep{li2024seist} introduced Seismogram Transformer (SeisT), a foundation model tailored explicitly for multiple seismic monitoring tasks, such as earthquake detection, seismic phase picking, and first-motion polarity classification. SeisT was trained on the DiTing dataset and evaluated for generalizability using the PNW dataset, both datasets containing comprehensive seismic events and associated annotations such as arrival times, magnitudes, and polarity information. These seismic foundation models (GeoFMs) have demonstrated significant potential across various seismological tasks, profoundly influencing research paradigms within seismology.

In the field of remote sensing, the work by Kuckreja et al. demonstrated the potential of the GeoChat model, developed by fine-tuning the LLaVA architecture and integrating a newly curated multimodal remote sensing dataset. This model can handle high-resolution remote sensing imagery and supports human-computer interaction through natural language, offering a novel toolset for interpreting and analyzing geophysical data~\cite{kuckreja2023geochat}. Additionally, \citep{liu2024remoteclip} proposed RemoteCLIP, the first vision-language foundation model dedicated to remote sensing. It leverages self-supervised learning and masked image modeling (MIM) strategies to generate robust visual features and semantic textual embeddings. To mitigate the issue of limited pre-training data, RemoteCLIP employs data augmentation techniques along with drone imagery, assembling a pre-training dataset that is 12 times larger than existing benchmarks. Experimental results indicate that RemoteCLIP surpasses traditional models across several tasks, achieving up to a 6.39\% improvement in accuracy over CLIP in zero-shot classification scenarios.

Several other foundation models integrating geophysical domain knowledge with language models, vision models, and generative models have been developed to better serve geophysical research. These include PetroQA~\cite{eckroth2023answering}, PetroBERT~\cite{rodrigues2022petrobert}, GeoGPT~\cite{zhang2024geogpt}, RockGPT~\cite{zheng2022rockgpt}, SAM-Oil~\cite{wu2024compositional}, RockSAM~\cite{ma2023zero}, and RockSE-Ferret~\cite{liu2024research}. These specialized foundation models significantly enhance the capabilities of geophysical interpretation, exploration, and analysis tasks.

These pioneering studies have demonstrated outstanding contributions in dataset construction, model development, and deployment, significantly advancing the adoption and application of foundation models within the geophysical domain. By employing meticulously designed data preprocessing strategies, task-specific optimizations, and cross-domain model innovations, researchers have effectively enhanced both the practical performance and scalability of foundation models in geophysical applications. Particularly, through the deep exploration of seismic and remote sensing data, as well as strong generalization across multiple tasks, these efforts have not only improved the accuracy of data analytics in geophysics but also laid a solid foundation for future research and technological development. Moving forward, our goal is to conceptualize and establish a comprehensive framework for constructing geophysical foundation models. Within this framework, we aim to explore strategies for developing more efficient, broadly applicable, and highly generalizable models, thereby driving sustained innovation and advancement in geophysics.

\section{Workflow and Considerations for Building Geophysical Foundation Models}
\label{sec:technology}

\begin{figure*}[!htb]%
\centering
\includegraphics[width=1\textwidth]{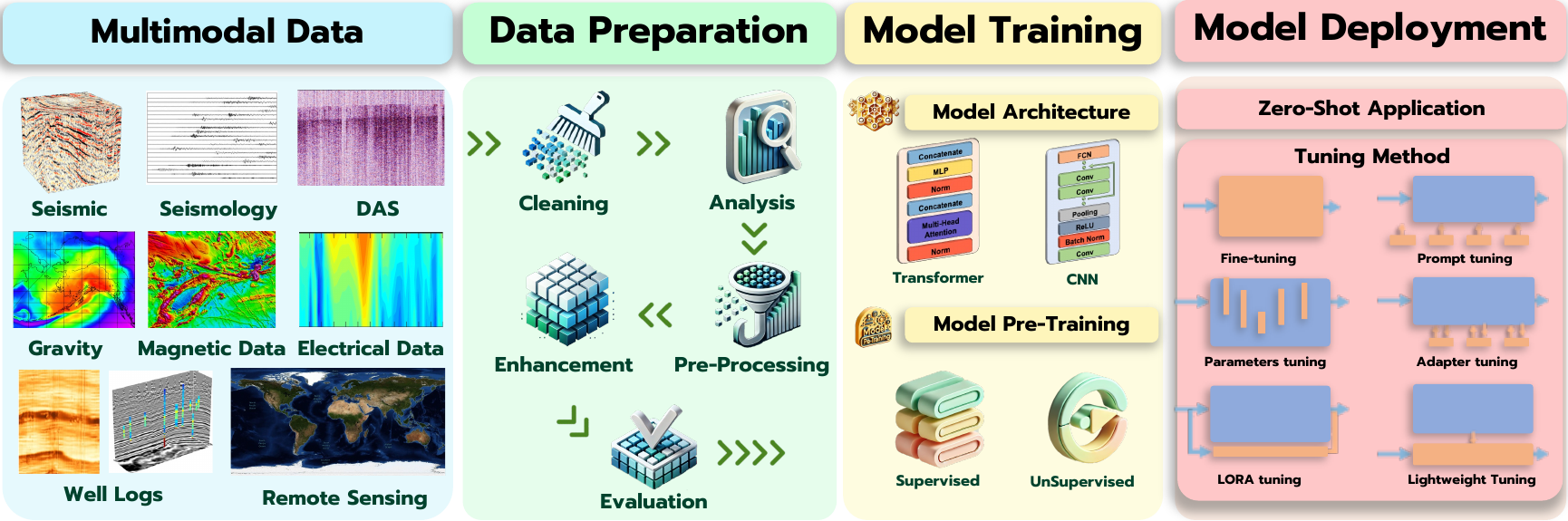}
\caption{The integration of foundation models into geophysical research involves several critical components, including data collection, preprocessing, model training, and model deployment. (Some images are modified from online sources.)
}
\label{fig:Fm3}
\end{figure*}

The development of foundation models involves multiple essential components, including feature diversification and data richness, rigorous data cleaning processes tailored to training requirements, the appropriate selection of model architectures, optimal pre-training methodologies, and effective deployment strategies (Figure~\ref{fig:Fm3}). In the following sections, we will systematically discuss and explore these key aspects, providing detailed insights into building efficient, robust, and widely applicable geophysical foundation models.

\subsection{Diversity of Features and Abundance of Geophysical Data}

\begin{figure*}[!htb]%
\centering
\includegraphics[width=1\textwidth]{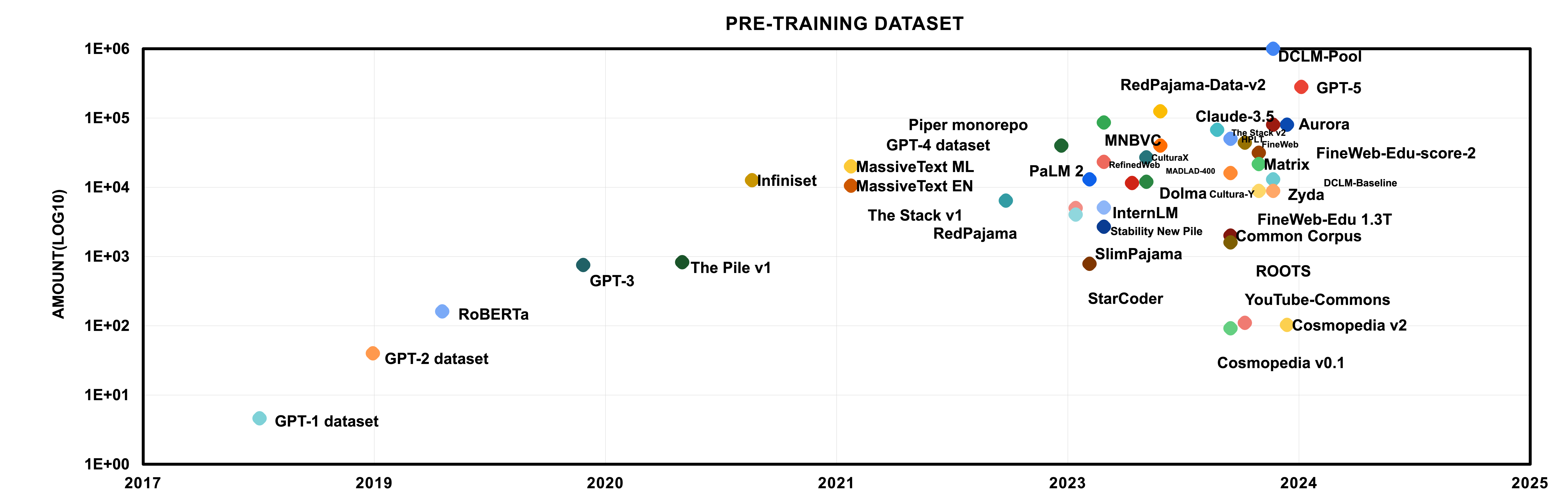}
\caption{The Growth Trend of Dataset Sizes for Training Large Language Models. The figure illustrates changes in the size of major pre-training datasets since September 2017, encompassing dataset scales such as GPT-1, GPT-2, GPT-3, and subsequent expansions into large-scale datasets. Over time, the dataset sizes have consistently increased, with a significant rise in both quantity and diversity observed especially from 2023 onward. This trend highlights the enormous amount of data required to train large language models and underscores the escalating computational demands involved~\cite{Thompson2024}.
}
\label{fig:datasets}
\end{figure*}

Data plays a pivotal role in the development of foundation models. Primarily, data constitutes the core material for training models; without high-quality data, a model cannot effectively learn or predict. Moreover, the diversity and richness of datasets directly impact the generalization capabilities of models, ensuring their adaptability across various scenarios. Additionally, robust data cleaning and preprocessing procedures enhance model accuracy and stability, mitigating the detrimental effects of noise. In essence, data is not merely the raw material for constructing foundation models—it fundamentally determines their performance and reliability. Furthermore, as illustrated in Figure~\ref{fig:datasets}, the scale of datasets used to train large language models has grown exponentially over time, alongside substantial increases in diversity. This indicates the necessity of simultaneously emphasizing both data scale and diversity when developing foundation models~\cite{Thompson2024}. Specifically in geophysics, large quantities of unlabeled data coexist with limited labeled datasets, both of which will be discussed in detail below.

\begin{figure*}[!htb]%
\centering
\includegraphics[width=1\textwidth]{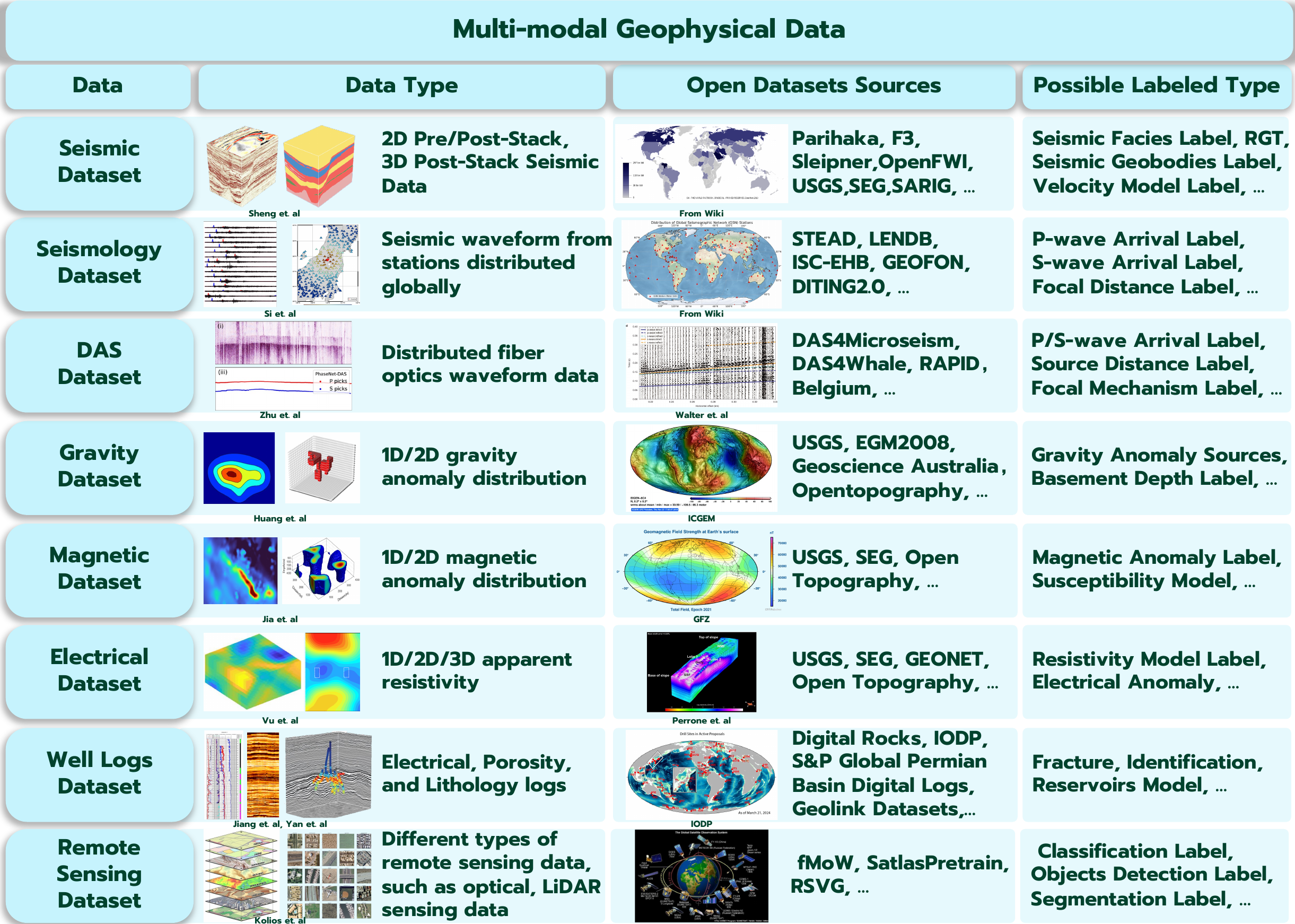}
\caption{Multimodal Datasets for Training Foundation Models in Geophysics. This figure presents various types of geophysical datasets along with their corresponding data types, sources of publicly available datasets, and potential annotation types. The principal datasets covered include seismic data, seismological data, distributed acoustic sensing (DAS) data, gravity data, magnetic data, electrical method data, well-logging data, and remote sensing data. Collectively, these datasets provide comprehensive data support for pre-training foundation models within the field of geophysics.
}
\label{fig:DataSource}
\end{figure*}

\subsubsection{Large-scale Unlabeled Geophysical Datasets}

In earth sciences, ongoing advancements in data acquisition, transmission, and storage, combined with continual improvements in data processing and interpretation methodologies, have led to an unprecedented accumulation of earth system observational data. However, geoscientific data is characterized by diverse sources and varied types, necessitating comprehensive preprocessing to convert such data into formats suitable for deep learning. To this end, extensive datasets have been collected and preprocessed by domain experts, standardizing data into unified formats. For instance, in exploration geophysics, due to confidentiality and commercial sensitivities, many exploration datasets are not openly available. However, numerous open-source unlabeled datasets still exist. Examples include datasets provided by the U.S. Geological Survey (USGS~\cite{USGS}), the South Australian Resources Information Gateway (SARIG~\cite{SARIG}), the Society of Exploration Geophysicists (SEG~\cite{SEG}), and the Netherlands dataset~\cite{Netherlands}, among others. These datasets encompass diverse geological characteristics, significantly benefiting the large-scale pre-training of foundation models. Log data is similarly accessible through multiple open resources, such as the International Ocean Discovery Program (IODP~\cite{IODP}), S\&P Global Permian Basin Digital Logs~\cite{SPGlobal}, and the Geolink database~\cite{Geolink}.

In seismology, extensive seismic data recorded by global permanent and temporary seismic stations are available, subject to institutional permissions. Similarly, in gravity, magnetic, and electromagnetic domains, numerous public data repositories exist, including USGS~\cite{USGS}, SEG~\cite{SEG}, Opentopography~\cite{Opentopography}, Geoscience Australia~\cite{GeoAustrialia}, EGM2008~\cite{pavlis2012development}, and GeoNet~\cite{GeoNet}. Additionally, several publicly accessible distributed acoustic sensing (DAS) datasets exist, such as passive-source DAS datasets (e.g., DAS4Microseism, DAS4Whale, RAPID, Belgium, Monterey Bay, SAFOD, FORESEE~\cite{spica2023pubdas}) and active-source DAS datasets (e.g., PoroTomo, FORGE 2C, Marcellus, Garner Valley, Levee Workshop~\cite{spica2023pubdas}).

\subsubsection{Labeled Geophysical Datasets}

Beyond large-scale unlabeled datasets, geophysicists have also dedicated efforts towards compiling labeled datasets suitable for supervised deep learning training. For instance, in exploration seismic applications, datasets such as the Parihaka seismic interpretation dataset, the F3 dataset for seismic facies classification, the TGS Salt dataset for salt dome segmentation, and the Sleipner dataset related to CO$_2$ injection detection~\cite{lin2024machine} have been widely utilized. Additionally, synthetically generated datasets have gained popularity for addressing practical problems. Examples include OpenFWI used as a benchmark for full waveform inversion imaging~\cite{deng2022openfwi}, FaultSeg3D for seismic fault interpretation~\cite{wu2019faultseg3d}, and the SEAM model for tasks such as salt dome segmentation, velocity, and density inversion. 

In seismology, various labeled seismic event and waveform datasets have been publicly released, including global-scale datasets such as STEAD, LENDB, ISC-EHB, and GEOFON. In addition, datasets addressing regional seismic events and waveforms—such as DITING2.0, DITING, SCEDC, Iquique, INSTANCE, and ETHZ—are available~\cite{arrowsmith2022big}. Furthermore, specialized datasets addressing volcanic eruptions, nuclear explosions, lightning strikes, and blast signals (e.g., PNW) have also been collected and processed.

In remote sensing, substantial labeled datasets have been developed, supporting visual tasks in geophysical analyses. Examples include the Functional Map of the World (fMoW) dataset, comprising over one million satellite images from more than 200 countries~\cite{fmow2018}. The SatlasPretrain dataset provides approximately 302 million labeled images across various categories, catering to diverse remote sensing tasks~\cite{bastani2023satlaspretrain}. Moreover, the Remote Sensing Vision and Grounding (RSVG) dataset contains numerous image-query pairs tailored for training multimodal remote sensing foundation models~\cite{zhan2023rsvg}. These labeled resources significantly enhance the capacity for deep learning and multimodal modeling, thus facilitating more effective analyses of geophysical data.

\subsubsection{Multimodal Geophysical Datasets}

In geophysical exploration, multiple data acquisition methods are frequently applied within the same region, involving diverse data types and collection techniques. These datasets can be regarded as multimodal, comprising different forms of information. Specifically, early geological surveys typically utilize textual and image data, providing geological descriptions and visualizations. Subsequent exploration phases employ various geophysical techniques such as gravity, magnetic, electrical, seismic, and logging methods, each acquiring distinct types of data to elucidate subsurface structures and resource distributions. For instance, gravity and magnetic surveys infer subsurface structures by detecting material properties, while electrical and seismic methods further analyze subsurface physical states and structural features. In later developmental stages, interpretative results regarding subsurface resources and geological evolution also constitute multimodal data. These datasets, collected through various methodologies, carry different physical meanings and spatial resolutions but can be integrated through data fusion techniques to establish a comprehensive subsurface model. Foundation models trained on these multimodal data can provide more accurate and integrated insights into subsurface structures and resource distributions.

Several public projects exemplify multimodal data availability in geosciences, such as the North Sea project, numerous CCUS initiatives (available at co2datashare.org), and the Volve project. These projects typically collect seismic, logging, microseismic, gravity, and electromagnetic data for identical regions, offering a robust foundation for developing multimodal models.

However, constructing multimodal deep learning models in geophysics presents several challenges. Firstly, data diversity complicates the fusion process, as seismic, gravity, magnetic, and electromagnetic data differ significantly in representation, resolution, and temporal scale, making effective alignment and fusion technically demanding. Additionally, multimodal geophysical models often require substantial computational resources, leading to bottlenecks during large-scale training and optimization. Although multimodal learning theoretically enhances model performance, its practical applications in geophysics remain limited. Traditional methods often rely on single-modal data, and introducing additional modalities may inadvertently introduce noise, negatively affecting performance. Typically, multimodal integration occurs during fine-tuning stages, posing efficiency challenges in integrating various data modalities. Future advancements in joint gravity-magnetic inversion and related techniques addressing data alignment and computational bottlenecks will likely bolster multimodal deep learning in geophysics, further advancing the field.

Geophysical datasets currently face challenges such as diverse formats for unlabeled data and insufficient labeled datasets. Unlabeled datasets exhibit considerable variability due to differences in research fields and measurement methods, lacking standardized formats and interfaces, complicating cross-disciplinary data sharing and analysis. Simultaneously, labeled datasets remain scarce, often task-specific, and reliant on expert annotation, resulting in slow production rates and limited diversity. To promote widespread data use, standardized data processing—including data cleaning, annotation and categorization, preprocessing, augmentation, and rigorous evaluation is urgently required.

\begin{figure*}[!htb]%
\centering
\includegraphics[width=1\textwidth]{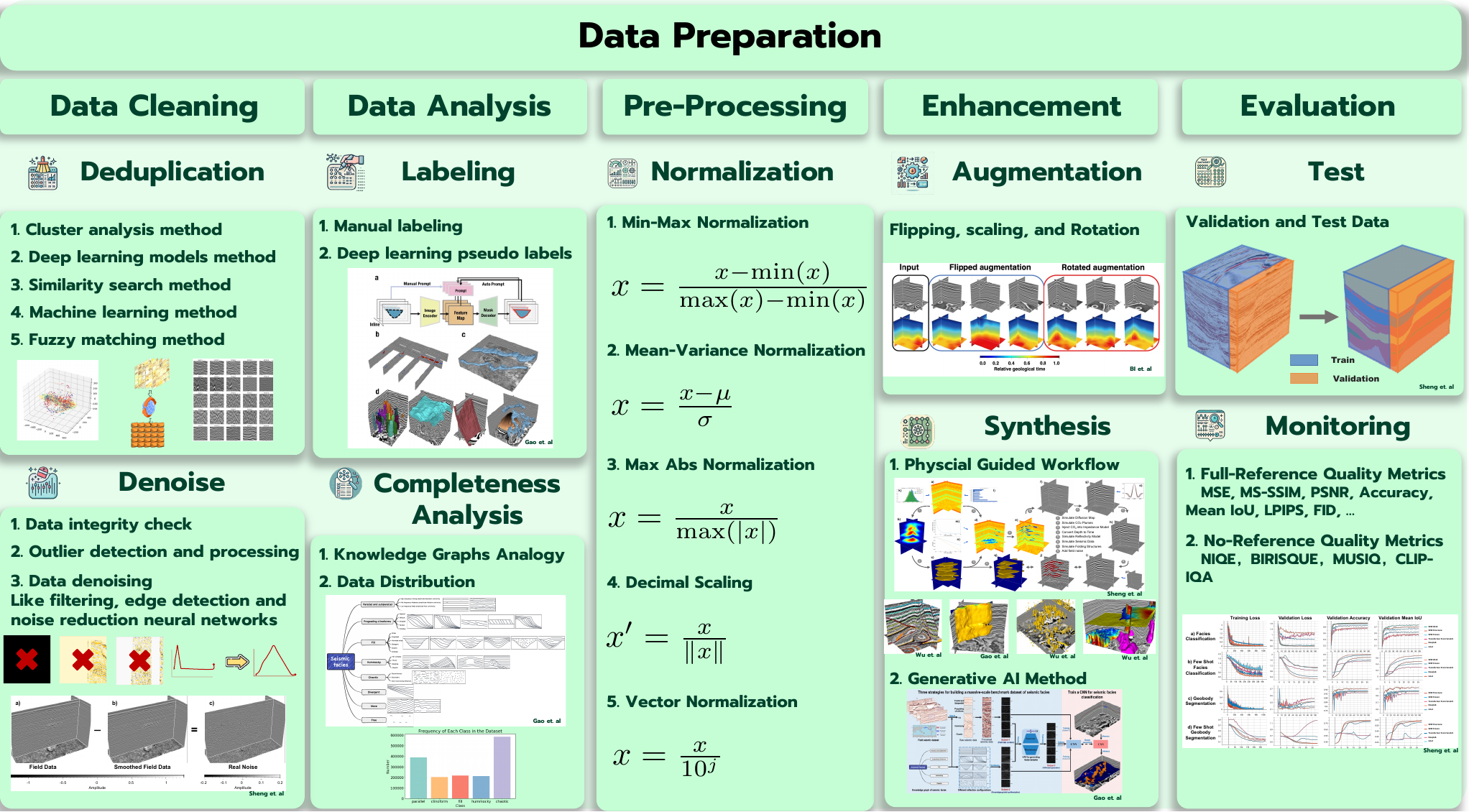}
\caption{The figure illustrates the various preprocessing stages geophysical data undergo before being used to train foundation models. The workflow encompasses data cleaning (e.g., deduplication and denoising), data analysis (e.g., labeling and completeness analysis), standardization (e.g., min-max normalization and mean-variance normalization), data augmentation (e.g., flipping, scaling, rotation), and the final preparation of validation and test datasets. The figure also demonstrates how physical-guided workflows can be integrated with generative AI methods for data synthesis, accompanied by continuous quality monitoring to ensure efficiency and accuracy in data preprocessing.}
\label{fig:DataPrePro}
\end{figure*}

\subsection{Preparation of Training Data for Foundation Models}
Preparing training data for foundation models is critical to ensuring model efficiency and accuracy, especially in geophysics. Carefully prepared, diverse, and high-quality data enable models to demonstrate robust generalization across varying geological conditions and environmental changes. Data cleaning, denoising, annotation, and augmentation enhance data quality, reduce noise interference, and facilitate effective fusion of multimodal data, thus providing comprehensive feature information for models. Such thorough data preparation allows models to better adapt to specific tasks and facilitates practical deployment and application. Consequently, data preparation decisively influences both model performance and practical effectiveness in geophysical applications.

\subsubsection{Data Cleaning}
Data cleaning is crucial for enhancing dataset quality and diversity. Deduplication and denoising are key components. Deduplication eliminates redundant data, avoiding overfitting and ensuring dataset diversity. Techniques include clustering methods (e.g., K-means~\cite{arthur2006k}, DBSCAN~\cite{ester1996density}), machine learning classifiers (e.g., Random Forest~\cite{breiman2001random}, SVM~\cite{cortes1995support}), NLP-based approaches (e.g., Word2Vec~\cite{mikolov2013efficient}, BERT~\cite{devlin2018bert}), fuzzy matching algorithms (e.g., Levenshtein distance, Jaccard similarity), deep learning models (CNNs~\cite{hinton2006reducing,he2016deep} , Transformers~\cite{vaswani2017attention}), specialized data-quality tools (Talend, Informatica), and distributed frameworks (Apache Hadoop, Spark).

Denoising removes data noise, such as corrupted, blank, or erroneous data, to enhance quality and facilitate learning. Methods include integrity checks (handling missing values), outlier detection (box plots, Z-score), consistency checks, machine learning models (isolation forest, LOF, autoencoders), NLP techniques for text denoising, image processing methods (filtering, edge detection, neural denoisers), and automated tools (OpenRefine, Pandas). These approaches effectively remove noise and redundancy, improving data quality and ensuring models learn meaningful information.
 
\subsubsection{Data Annotation and Completeness Analysis}
  
It is crucial to annotate and classify data appropriately based on the type of model and its training requirements. Data annotation and categorization are vital steps in data processing and model training, enhancing data management efficiency and providing precise data inputs to the models. For supervised learning models, detailed data annotation is essential; for example, in image classification tasks, each image must be labeled with its corresponding class. Although unsupervised or self-supervised learning models do not require large-scale labeled data, they still need certain preprocessing steps to facilitate model learning.

Ensuring the completeness of geophysical data is essential for guaranteeing data quality and usability. Geophysical data can be classified into seismic data, remote sensing imagery, seismic waveform data, and geological profiles, each serving specific geological research fields. Each data type requires verification of its completeness. For example, seismic data should include both pre-stack and post-stack datasets, ensuring critical information is not omitted during seismic exploration and analysis. Completeness analysis can be enhanced through knowledge graph construction methodologies, leveraging geological domain knowledge to evaluate the completeness comprehensively, ensuring the datasets meet scientific research needs. Category statistics based on knowledge graphs and predefined categories can effectively detect biases within the data. Knowledge graphs facilitate understanding inter-category relationships and distributions, assisting in identifying missing or imbalanced categories, thus enabling adjustments via oversampling or undersampling. Classification-based approaches, such as decision trees, Support Vector Machines (SVM), and neural networks, can evaluate data distributions to identify biases or imbalances timely. Statistical methods, including Chi-square tests, Kolmogorov-Smirnov (K-S) tests, and maximum likelihood estimation (MLE), further verify category completeness statistically, mitigating potential biases. Additionally, strategies such as multi-source data fusion, data augmentation, and weighted loss functions effectively prevent biases arising from data deficiency or category imbalance, ensuring fairness and representativeness of datasets.

Geophysical data can be further organized based on geological environments—categorized by depositional models, structural models, geographic feature distribution patterns, and research objectives. For example, hydrological data support studies of rivers, lakes, and groundwater dynamics, while volcanic monitoring data contribute to volcanic activity research. Temporal data can be structured chronologically, including historical earthquake records, real-time monitoring datasets, and climate change data. Moreover, data attributes can differentiate structured, unstructured, and semi-structured data, enhancing processing efficiency and precision. Through comprehensive completeness checks, the quality of datasets is ensured, enhancing foundation model training outcomes and providing a robust data foundation for geophysical research.

When conducting completeness analysis of geophysical data, category statistics leveraging knowledge graphs and predefined categories effectively identify biases. Knowledge graphs analyze relationships and distributions between categories, aiding in detecting missing or imbalanced classes, which can subsequently be balanced through oversampling or undersampling methods. Classification model-based methods, such as decision trees, Support Vector Machines (SVM), and neural networks, evaluate category distributions through known classes, promptly identifying imbalance or biases. Statistical methods, including Chi-square tests, K-S tests, and maximum likelihood estimation (MLE), statistically validate category completeness, reducing potential biases. Additionally, employing multi-source data fusion, data augmentation, and weighted loss functions mitigates biases stemming from data deficiencies or category imbalance, ensuring datasets are fair and representative. These classification methods facilitate efficient organization and management of geoscientific data, ensuring effective data utilization, providing targeted model inputs, and thereby enhancing model performance and accuracy.

\subsubsection{Data Standardization}
Due to the high diversity in acquisition and processing of geophysical data, standardization preprocessing methods, such as mean-variance normalization and min-max normalization, are typically required before model training to mitigate inconsistencies. Geophysical data are acquired using diverse instruments, including seismometers, gravimeters, and magnetometers, whose varying sensitivities and operational principles may cause disparities in data quality, formats, and scales. Differences in acquisition methodologies—such as grid sampling, sampling rates, amplitude distributions, and spectral characteristics—further amplify data heterogeneity, affecting model training effectiveness. For instance, seismic data sampling rates and spatial resolution can vary significantly across instruments, potentially leading to loss or inaccuracies of detailed information in specific regions.

Given this context, temporal-spatial standardization becomes indispensable. Geophysical data exhibit strong temporal and spatial dependencies, often showing significant differences across various temporal-spatial scales. Temporal-spatial corrections unify data within a standardized framework, ensuring consistency. These standardized preprocessing procedures effectively mitigate inconsistencies caused by varying acquisition conditions, minimizing related errors, thus enhancing the accuracy and generalization capabilities of trained models.

\subsubsection{Data Augmentation}
Data augmentation enhances a model’s generalization capability by generating diverse data samples. Augmentation strategies generally fall into two categories: data expansion and synthetic data generation. Data expansion involves applying transformations to existing data—for instance, flipping, scaling, or rotating images—to create a more varied set of training samples, thereby improving the robustness of the model.

Synthetic data generation, on the other hand, is employed when there is a need to increase the size and diversity of the dataset by producing entirely new data samples. Common approaches include using physics-based simulation methods or generative models such as Generative Adversarial Networks (GANs)\cite{goodfellow2014generative} and Diffusion Models\cite{rombach2022high}, which can generate new image samples to enrich the training dataset and enhance model performance.

In geophysics, generating data for neural network training often relies on the integration of physical equations and numerical simulation techniques. Physical models—such as wave equations, heat conduction equations, or electromagnetic equations—can simulate subsurface and surface geophysical processes, thereby producing high-quality training data. These equations provide accurate representations of geophysical phenomena and can be used to simulate seismic wave propagation, electromagnetic field distributions, and thermal variations under different conditions. Specifically, numerical simulation methods such as the Finite Element Method (FEM), Finite Difference Method (FDM), and Finite Volume Method (FVM) are used to solve these equations, generating simulated data under varying geological backgrounds, terrains, and physical conditions. The results of these simulations can then be used to construct physically meaningful input datasets for training neural networks. For example, seismic data can be generated by simulating the propagation of elastic waves through the subsurface, while electromagnetic data can be obtained by solving Maxwell's equations for electric and magnetic fields. This physics-based data generation approach offers several advantages. It provides rich physical context and ensures data authenticity and reliability. Moreover, it addresses the challenge of limited availability of real-world exploration data, enabling the construction of sufficiently large datasets required for training deep learning models. This, in turn, enhances model generalization and prediction accuracy~\cite{sheng2023deep}. Furthermore, by simulating diverse physical scenarios, this method enables the creation of heterogeneous training data, improving the model’s adaptability to complex geological environments. 

However, it is important to note that synthetic data may differ from real exploration data in certain respects. As a result, data augmentation and domain adaptation techniques are typically employed during training to bridge the domain gap between simulated and real-world data, thereby further improving model performance. In geophysics, neural network-based approaches—particularly Generative Adversarial Networks (GANs) and Diffusion Models—are emerging as powerful tools for data generation and augmentation. These methods not only enable the creation of high-quality synthetic geophysical data in data-scarce scenarios but also enhance model accuracy and generalization by learning the underlying physical patterns of geophysical processes. A GAN consists of two components: a generator and a discriminator. The generator is responsible for producing synthetic geophysical data (e.g., seismic wave propagation, resistivity distributions), while the discriminator evaluates the similarity between the generated data and real data. During training, the generator iteratively improves its outputs to the point where the discriminator can no longer distinguish between real and generated samples. In this way, GANs can enrich training datasets with simulated geophysical data, thereby facilitating the training of deep learning models and improving their performance in real-world exploration scenarios. \citep{puzyrev2022geophysical} proposed a deep generative framework for geophysical inversion by leveraging Generative Adversarial Networks (GANs) to synthesize high-resolution 2D subsurface models. By incorporating stratigraphic modeling tools such as Badlands to generate realistic geological scenarios, and training networks separately on density and stratigraphy distributions, the framework efficiently produces diverse, geologically plausible samples in real time. This approach addresses the critical bottleneck of limited training data in data-driven inversion workflows and enables the construction of large-scale synthetic datasets in a cost-effective manner, thereby enhancing the robustness and generalizability of deep learning models for seismic, gravity, and electromagnetic inversion tasks. The work offers a scalable solution to bridge data scarcity and model training efficiency in geophysical interpretation.

In contrast, Diffusion Models generate data by gradually introducing noise to real data and then learning to reverse this process through denoising steps during training. These models have demonstrated impressive capabilities in image generation and show strong potential in geophysical data synthesis as well. By injecting noise into seismic or electromagnetic data and then reconstructing it through a learned denoising process, diffusion models can simulate realistic geophysical scenarios, capturing complex geological conditions and physical phenomena. Compared to GANs, diffusion models often yield more stable generation results and produce more diverse, physically consistent data. \citep{wang2024controllable} introduced a conditional generative diffusion framework for seismic velocity model synthesis, aiming to overcome limitations in data availability and prior incorporation in machine learning-based inversion. By integrating diverse prior information—including geological classes, well logs, and reflectivity images—into the diffusion process, their method enables the generation of velocity models that are both geologically plausible and tailored to specific inversion scenarios. Trained on the OpenFWI dataset, the model demonstrates strong generalization, even under out-of-distribution conditions, highlighting its potential to produce high-fidelity, prior-informed datasets for training robust velocity inversion networks. This work provides a flexible and scalable pathway for enriching data-driven geophysical inversion with expert knowledge and multimodal constraints. The application of both GANs and diffusion models makes the geophysical data generation process more flexible and versatile. By learning the latent distribution of geophysical datasets, these models can generate physically plausible data even in the absence of sufficient real-world observations. The synthetic data produced can not only enhance model training but also support tasks such as data augmentation, denoising, and geophysical inversion, thereby improving data quality and analytical precision in geophysical exploration. 

\subsubsection{Evaluation and Testing}
  
To ensure the effectiveness of the model, it is essential to prepare an independent test dataset and continuously monitor performance throughout the training process. The test dataset should be disjoint from the training dataset and representative of real-world data distributions. This allows for an accurate assessment of the model’s performance on previously unseen data and provides a reliable measure of its generalization capability.

During training, it is also critical to continuously monitor both data quality and model performance, with particular attention to training and validation errors. This facilitates early detection of overfitting or underfitting issues and enables timely adjustments. Data preparation and model training are iterative optimization processes, requiring careful attention to several key aspects. First, potential biases in the dataset must be identified and mitigated to ensure model fairness and prevent inaccurate predictions due to underrepresented categories. Second, improving model interpretability through feature selection techniques to identify key predictive features is crucial for enhancing model transparency. Lastly, data preparation should be treated as a dynamic process: one that evolves based on model performance and experimental requirements. This may involve augmenting the training dataset or revisiting data cleaning and annotation procedures to progressively improve data quality and model outcomes.

Through meticulous data preparation and preprocessing, it is possible not only to improve the learning efficiency of the model but also to enhance its adaptability to complex geological environments. The data preparation pipeline plays a vital role in eliminating irrelevant noise, ensuring data consistency and quality, and providing a solid foundation for stable and reliable neural network training.

\subsection{Model Architecture for Foundation Models}
Once data preparation is complete, the next critical step is the design of an appropriate model architecture and the selection of a suitable pretraining strategy. The architecture should be tailored to the specific task requirements and data characteristics to ensure that the model can effectively capture the underlying patterns in the data. The pretraining strategy involves selecting appropriate proxy tasks and training on large volumes of relevant data, often unlabeled, to enable the model to extract meaningful geophysical features. By aligning the proxy tasks with the practical needs of geophysical applications, the model can be endowed with the capacity to accurately and efficiently learn domain-relevant representations.

\begin{figure*}[!htb]%
\centering
\includegraphics[width=1\textwidth]{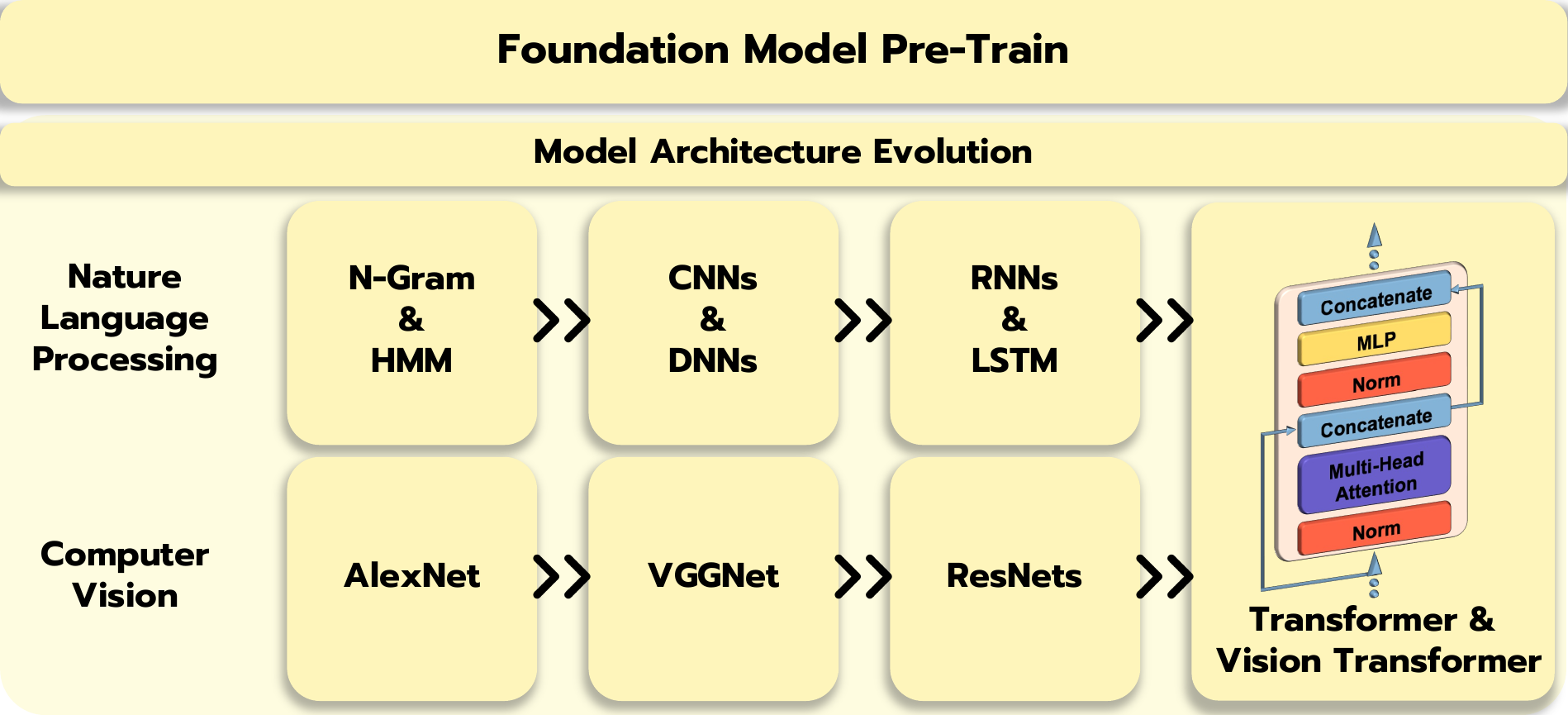}
\caption{ This figure illustrates the evolution of model architectures across natural language processing, computer vision, and multimodal tasks—from early models such as N-Grams and Hidden Markov Models (HMMs), to convolutional neural networks (CNNs), recurrent neural networks (RNNs), and long short-term memory (LSTM) networks, culminating in the modern Transformer and Vision Transformer architectures. Regarding pretraining strategies, the figure outlines both supervised and unsupervised learning approaches, including masked token prediction, autoregressive objectives, generative methods, and contrastive learning paradigms. For multimodal tasks, it also highlights text-to-image generative pretraining strategies.}
\label{fig:Model}
\end{figure*}

\subsubsection{Advancements and Advantages of the Transformer Architecture}

The predominant architecture employed in foundation models today is the Transformer architecture (see Fig.\ref{fig:Model}). The Transformer was first introduced by Vaswani et al. in 2017\cite{vaswani2017attention}, marking a major breakthrough in the field of natural language processing (NLP). Unlike traditional recurrent neural networks (RNNs)\cite{mikolov2010recurrent} and long short-term memory networks (LSTMs)\cite{gers2000learning}, the Transformer is entirely based on the self-attention mechanism, discarding recurrent structures. This allows for fully parallel computation, significantly improving training efficiency while simultaneously processing all elements in a sequence.

In recent years, the architecture of vision models has also moved towards unification, with the Transformer playing a dominant role. Leveraging its self-attention mechanism, the Transformer excels at capturing long-range dependencies within images, overcoming the limitations of convolutional neural networks (CNNs) that primarily focus on local feature extraction. With the emergence of models such as the Vision Transformer (ViT)~\cite{dosovitskiy2020image}, the Transformer architecture has not only enhanced the ability to understand visual content but has also demonstrated increased flexibility in handling multimodal data. As a result, an increasing number of vision models have adopted Transformer-based architectures, driving the field toward greater architectural consistency and computational efficiency.

The Transformer offers several key advantages, including powerful parallel computing capabilities, the ability to model long-range dependencies, excellent scalability, and effective multimodal fusion. Its ability to process all elements in a sequence simultaneously leads to significant improvements in both training and inference efficiency. Moreover, it excels at handling tasks requiring complex contextual understanding, particularly those involving long-range interactions. The Transformer is also highly adaptable across a range of data modalities—including text, images, and more—making it particularly well-suited for applications in geophysics, medicine, and remote sensing. Owing to these strengths, the Transformer has demonstrated outstanding performance in large-scale datasets and complex tasks across a wide range of domains.

In the field of geophysics, the Transformer architecture has demonstrated significant potential in handling spatiotemporal, temporal, and spatial data. Given the unique characteristics of geophysical datasets, it is important to examine the adaptability of Transformer-based models and their specific implementation strategies. Temporal sequences—such as seismic wave propagation, climate variation records, and other time-dependent measurements—are common in geophysical studies. The Transformer has shown strong performance in modeling time series data, particularly due to its capacity to capture long-range dependencies. Its self-attention mechanism allows the model to learn complex interdependencies across different time steps, which is especially valuable in fields such as seismology and meteorology. By enabling the transmission of information between arbitrary positions in the sequence, the Transformer effectively overcomes the vanishing gradient problem often encountered in traditional recurrent neural network (RNN) architectures. Nevertheless, geophysical time series data may include noise or missing values, which poses challenges for Transformer-based models. To address this, noise suppression and missing value imputation techniques can be applied during data preprocessing. Alternatively, robust training strategies can be incorporated into the model to enhance its resilience to imperfect real-world data. Spatial data in geophysics often take the form of gridded or three-dimensional geospatial datasets, which resemble image and video data in computer vision. The Transformer’s self-attention mechanism is also well-suited to modeling complex correlations across both temporal and spatial dimensions in such spatiotemporal datasets. To capture these relationships effectively, modified Transformer architectures can be designed. For instance, convolutional operations can be integrated into the spatial dimension to extract local features, similar to the convolutional layers introduced in the Vision Transformer (ViT). Additionally, specialized space-time attention mechanisms can be developed to optimize performance on spatiotemporal geophysical data.

While convolutional neural networks (CNNs) have inherent advantages in extracting local spatial features, the Transformer excels at capturing global dependencies through its self-attention mechanism. A hybrid architecture that combines the strengths of both CNNs and Transformers can therefore be particularly effective for spatial data in geophysics. For example, CNNs can be used initially to extract spatial features, which are then passed to a Transformer to model global spatiotemporal relationships. To further improve performance, spatial data can be partitioned into smaller patches—analogous to image patches in vision models—with each patch treated as a token input to the Transformer. This approach preserves local spatial detail while leveraging self-attention to learn global dependencies, making it well-suited for the complex spatial structures commonly encountered in geophysical applications.

\subsubsection{Adapting Transformers to Geophysical Data}

 \begin{itemize}
        \item \textbf{Input Layer Adaptation}
 Geophysical data should first be transformed into a format suitable for processing by Transformer models. Geophysical datasets  exhibit clear multimodal characteristics, involving dimensions such as time, space, frequency, and imagery. To enable foundation models to effectively understand, process, and extract valuable information from these continuous and complex datasets, they are typically converted into discrete units called "tokens," a process known as tokenization.

In practical implementations, tokenization methods generally fall into two categories: hard tokenization and soft tokenization. Hard tokenization explicitly partitions data into discrete segments with clearly defined boundaries and assigned identifiers. For instance, seismic waveforms can be segmented into fixed one-second intervals, and images can be divided into regularly spaced grid patches. This method provides clarity and stability, making it particularly suitable for tasks that require efficient processing and rapid indexing. In contrast, soft tokenization encodes data into continuous feature vectors or embeddings without strictly defined boundaries, allowing tokens to flexibly capture complex semantic interactions. These soft tokens are typically generated using neural network methods: Convolutional Neural Networks (CNNs) produce continuous feature maps encoding local visual information; self-attention mechanisms within Transformers capture long-range contextual relationships; diffusion models propagate spatial information continuously across images; and autoencoders, such as Variational Autoencoders (VAEs), compress data into concise latent embeddings. 

For time-series geophysical data, such as seismic waveforms or electromagnetic responses, hard tokenization typically involves segmenting the waveform into uniform intervals and encoding each interval into discrete tokens via methods like vector quantization (e.g., VQ-VAE). Conversely, soft tokenization relies on self-supervised approaches such as HuBERT or convolutional networks, producing continuous contextualized vector representations. Both approaches have demonstrated success in applications such as seismic event recognition, magnitude prediction, and waveform clustering, each with its own strengths.

In the context of image-based geophysical data like seismic profiles, gravity anomaly maps, or resistivity distributions, hard tokenization commonly divides images into fixed-sized visual patches (e.g., 16×16 pixels), mapping each patch into an individual visual token. Soft tokenization, on the other hand, utilizes deep neural networks to automatically learn continuous image features, embedding these into a high-dimensional semantic space. For example, the CLIP model can map image patches into a shared semantic space with text, enabling sophisticated image-text interactions. This approach is particularly advantageous for precisely representing complex geological structures or ambiguous boundaries.

For spatio-temporal data, such as time-varying resistivity profiles or seismic activity distribution maps, tokenization involves integrating both spatial and temporal dimensions. Hard tokenization typically segments data into structured spatiotemporal cubes (e.g., [x, y, t]), providing clear sampling structures and granularity control. Soft tokenization, however, permits each token to represent continuous similarity weights across multiple spatial locations and time points, effectively capturing continuous geological evolution. This makes soft tokenization highly suitable for modeling phenomena such as seismic migration and fault growth dynamics.

From an application perspective, hard tokenization for seismic waveforms offers efficiency and easy indexing, whereas soft tokenization better captures nuanced semantic information. In seismic imaging tasks, hard tokenization simplifies visualization and labeling due to its structured nature, whereas soft tokenization excels at accurately capturing fuzzy or complex geological features. For spatiotemporal modeling, hard tokenization facilitates controlled granularity and regular sampling, while soft tokenization is advantageous in representing continuously evolving geological processes.

Overall, the tokenization of geophysical data is becoming increasingly systematic, combining the structural clarity of hard tokenization with the semantic continuity of soft tokenization. This hybrid approach enhances both computational efficiency and representational power, enabling artificial intelligence to more effectively interpret the complex "language" of the Earth's interior. Consequently, this integrated method provides powerful tools for addressing critical geophysical challenges, including earthquake prediction, resource exploration, and subsurface structural interpretation.

\item \textbf{Adaptation of the Self-Attention Mechanism}
While the self-attention mechanism excels at capturing long-range dependencies, its computational complexity can become a bottleneck when applied to large-scale spatiotemporal data. To address this challenge, hierarchical or multi-scale attention mechanisms can be introduced, or the temporal and spatial dimensions can be modeled separately. These strategies help reduce computational overhead and improve model efficiency while preserving the attention mechanism’s capability to capture relevant dependencies.

\item \textbf{Adaptation of Positional Encoding}
One of the core strengths of the Transformer architecture lies in its ability to encode positional information within sequences. For geophysical data, both temporal and spatial position information must be encoded appropriately. Temporal positional encoding can be adapted directly from established methods used in natural language processing. However, spatial positional encoding requires adjustments to accommodate two-dimensional or three-dimensional structures, especially in the context of irregular grids or discretized coordinates common in geophysical datasets.

Traditional Transformers typically rely on regular 1D or 2D positional encoding to capture structural information, which may be insufficient for geophysical data distributed on non-uniform grids. To address this, customized encoding schemes can be designed based on the spatial characteristics of the data, preserving both geographical and physical semantics. Such tailored encodings not only enhance the model’s understanding of irregular spatial layouts but also improve its accuracy and efficiency in handling complex geophysical tasks.

\item \textbf{Development of Mixture-of-Experts (MoE) Systems for Geophysical Data}
Mixture-of-Experts (MoE) architectures can enhance model performance by incorporating multiple specialized expert models, each optimized for particular data domains or tasks. In geophysics, data complexity and diversity arise from heterogeneous observation sources, various types of physical phenomena, and multi-scale spatial structures. MoE models allow individual experts to focus on specific data characteristics or problem domains. For instance, dedicated experts can be developed for different geophysical modalities such as gravity, magnetic, and electromagnetic data. These specialized modules can then be integrated to produce more accurate and comprehensive analytical results.

A gating network is typically used in MoE systems to determine which experts should process each input sample, thereby reducing computational cost by avoiding the activation of all experts simultaneously. The intrinsic heterogeneity of geophysical data makes MoE architectures particularly suitable for flexible and task-specific expert selection. This capability is especially beneficial for solving complex geophysical problems such as subsurface imaging, exploration data interpretation, and seismic wave propagation modeling. By leveraging domain-specific expertise and computational efficiency, MoE models improve the expressiveness and scalability of deep learning frameworks in large-scale geophysical data processing.

 \end{itemize}

\subsection{Pretraining Strategies for Foundation Models}
Pretraining strategies for foundation models can be broadly categorized into supervised and self-supervised approaches. These strategies fundamentally influence how the model learns data representations and determine its generalization ability across downstream tasks. Supervised pretraining typically relies on manually labeled datasets, where explicit supervision signals guide the model to extract features and recognize patterns. In contrast, self-supervised pretraining leverages the intrinsic structure and statistical regularities within the data to construct pseudo-supervised tasks, allowing the model to learn rich representations and contextual information from large volumes of unlabeled data.

The distinction between these strategies has significant implications for the resource requirements and data dependencies during pretraining, and directly affects the model’s generalizability and adaptability in downstream applications. In the following sections, we first introduce supervised pretraining strategies, followed by an in-depth discussion of self-supervised methods, covering recent advances in self-supervised language models, visual models, and multimodal pretraining techniques.

\begin{figure*}[!htb]%
\centering
\includegraphics[width=1\textwidth]{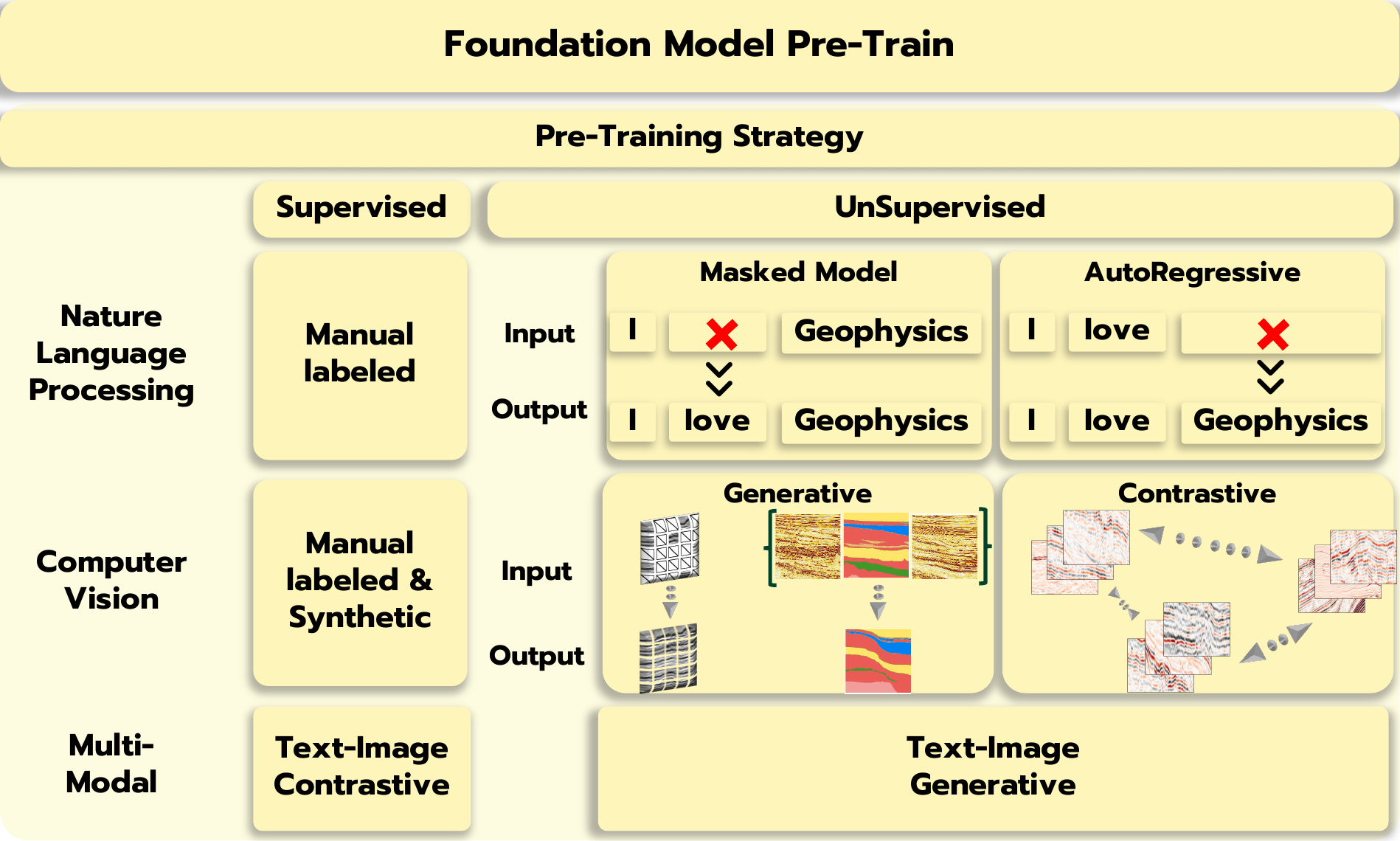}
\caption{Pretraining strategies for geophysical foundation models. The figure illustrates different approaches to supervised and self-supervised learning, including masked token prediction, autoregressive modeling, generative and contrastive pretraining strategies. For multimodal tasks, the figure also presents text-to-image generative pretraining techniques.
}
\label{fig:Training}
\end{figure*}

\subsubsection{Supervised Pretraining of Foundation Models}
Foundation models have achieved remarkable success across multiple domains, including natural language processing and computer vision, where different pretraining strategies have played a pivotal role (see Fig.\ref{fig:Training}). In current supervised learning paradigms, pretraining foundation models typically requires access to large-scale annotated datasets. For example, the Segment Anything Model (SAM)\cite{kirillov2023segment} was pretrained through large-scale supervised learning on an extensive dataset. According to publicly available information, SAM was trained on over one billion labeled data samples, encompassing a broad range of scenes and objects. This vast amount of supervision enables SAM to generalize effectively across various image segmentation tasks, even without task-specific fine-tuning.

Training such models necessitates the development of scalable data annotation pipelines. These often involve a combination of manual annotation, machine-assisted labeling, human verification, and fully automated labeling systems to construct high-quality, large-scale datasets. Despite automation efforts, data annotation remains a labor-intensive and resource-demanding process.

In the context of geophysics, the integration of foundation models is emerging as a frontier in exploration geophysics and seismology. In seismology, historical archives and extensive earthquake event recordings have led to the availability of modestly sized labeled datasets. However, these datasets are often collected by different seismic networks or institutions, resulting in heterogeneity in data formats and quality. This introduces challenges in terms of data standardization, cleaning, and consolidation. Moreover, while the overall data volume may be large, labeled seismic events of interest—particularly anomalous events—are often sparse and unevenly distributed. This imbalance necessitates careful sample selection and training strategies to avoid biased learning and ensure robust generalization.

By contrast, in exploration geophysics, large-scale labeled datasets are significantly more limited. Traditional datasets, including seismic, well-logging, electromagnetic, gravity, and magnetic data, are often unannotated or lack systematic labeling, and manual annotation is costly and time-consuming. Therefore, self-supervised and unsupervised pretraining strategies offer clear advantages in this domain. By leveraging the inherent physical laws and spatial structures within the data, it is possible to design effective pseudo-supervised tasks that enable foundation models to learn generalizable representations of subsurface geological structures and petrophysical properties from vast amounts of unlabeled data. This approach enhances the robustness and transferability of representations for downstream geophysical exploration tasks.

As the volume of data continues to grow, self-supervised learning has become the predominant approach for model pretraining. Self-supervised pretraining includes both language-model-based and vision-model-based strategies, which extract inherent features and patterns from the data through proxy tasks. These strategies provide generalized feature representations for downstream tasks in each modality. In addition to single-modality approaches, multimodal joint pretraining enables the integration of information across multiple data types. This method involves designing joint training objectives that allow the model to learn interactions and correlations between modalities such as image-text or audio-visual pairs. Consequently, multimodal pretraining supports cross-modal understanding, generation, and retrieval tasks, and is well-suited for complex scenarios involving heterogeneous data sources.

The continuous evolution and innovation of foundation model pretraining strategies have significantly advanced the application of artificial intelligence across a wide range of domains. Each type of pretraining approach offers distinct advantages and is suitable for different task scenarios. From single-modality self-supervised pretraining to contrastive learning and multimodal joint pretraining, pretraining technologies are rapidly evolving to meet the increasing complexity of real-world AI applications. Looking ahead, with the continued expansion of data availability and advances in computational resources, foundation model pretraining is expected to achieve further breakthroughs, opening new possibilities for the development of intelligent systems.

\subsubsection{Self-Supervised Language Model Pretraining}

Self-supervised language model pretraining is primarily conducted using large-scale text corpora. Models learn statistical regularities and semantic representations of language by performing tasks such as predicting masked tokens (as in Masked Language Models, or MLMs) or predicting the next word in a sequence (as in Auto-Regressive Language Models). For example, in the BERT model~\cite{devlin2018bert}, input text is randomly masked, and the model is trained to predict the masked tokens based on the surrounding context. In contrast, GPT models~\cite{openai2023gpt4} use auto-regressive objectives to predict the next token given the preceding context, enabling capabilities such as zero-shot learning.

Self-supervised language pretraining effectively captures contextual information, allowing the model to acquire rich semantic and syntactic knowledge. Training on massive corpora leads to highly generalizable textual representations. These pretrained models can be widely applied to a variety of natural language processing (NLP) tasks, such as text classification, sentiment analysis, and question answering. For instance, in text classification tasks, the pretrained language model can be used as a feature extractor, followed by a classification head to improve predictive accuracy.

In addition, multilingual language model pretraining involves using textual data in multiple languages. These models learn both the commonalities and differences across languages, often sharing model parameters to facilitate cross-lingual knowledge transfer. For example, mBERT (multilingual BERT) is pretrained on Wikipedia data from multiple languages and employs a unified model architecture to process texts across languages. This multilingual pretraining approach leverages linguistic diversity to enhance the model’s understanding of language universals. It also enables effective cross-lingual performance with limited training data in the target language—for example, in tasks like cross-lingual text classification or machine translation.

Such multilingual models are well-suited to multilingual information processing scenarios, including corporate document management in multinational enterprises and multilingual social media analysis. In machine translation tasks, multilingual pretrained models can first encode the source language and then generate the target language, reducing dependence on large-scale parallel corpora.

In the field of geophysics, training self-supervised language models begins with the collection of domain-specific textual data, such as academic publications, technical reports, textbooks, and related datasets. These sources contain a wealth of specialized terminology and technical details, forming the foundational corpus for model training. Designing appropriate pretraining tasks—such as Masked Language Modeling (MLM) and Auto-Regressive Language Modeling (ARLM)—is crucial, as these tasks enable the model to learn the semantic relationships between concepts and terms specific to geophysics.

When selecting model architectures, Transformer-based models such as BERT, GPT, CLIP, or T5 can be adapted to meet the needs of geophysical text processing. The training process typically consists of two stages: pretraining and fine-tuning. In the pretraining stage, the model acquires a general understanding of the linguistic patterns in geophysical texts. During fine-tuning, the model is optimized for specific downstream tasks, such as seismic data interpretation or mineral deposit prediction.

Model evaluation should not only focus on performance metrics such as accuracy on task-specific benchmarks but also consider aspects like interpretability and multimodal integration capabilities. These criteria ensure that the model can provide valuable insights and guidance for addressing geophysical problems. To maintain model relevance and effectiveness over time, it is essential to periodically update the training corpus and conduct incremental training to incorporate the latest advancements and knowledge in the geophysical sciences.

\subsubsection{Self-Supervised Pretraining for Visual Models}
Self-supervised pretraining for visual feature learning is typically based on proxy tasks derived from image data. Examples include predicting the rotation angle of an image (Image Rotation Prediction), estimating the relative positions of image patches (Relative Patch Location Prediction), or reconstructing masked regions of an image (Masked AutoEncoder, MAE~\cite{He_2022_CVPR}). These tasks allow the model to learn meaningful visual representations without requiring manual annotations.

For instance, a foundation model can apply a pretraining task in which shuffled image patches must be reordered into their original configuration. This forces the model to capture spatial structure and semantic content. The MAE method, specifically, trains the model to reconstruct the missing portions of an image after a significant portion of its pixels have been masked out~\cite{He_2022_CVPR}. In this approach, a large fraction of the image is occluded—set to zero or otherwise masked—and the model is trained to infer the missing content from the visible portions. These techniques significantly reduce the reliance on labeled data while enabling the model to extract valuable visual features directly from raw images. The visual features obtained through such self-supervised pretraining are highly generalizable and can be transferred to a wide range of downstream vision tasks, such as object detection, image segmentation, and scene understanding. Bai et al.~\cite{bai2024sequential} proposed a novel sequence modeling approach that represents raw images, videos, and annotated data using “visual sentences.” This method enables the training of large-scale vision models without relying on language data. By minimizing cross-entropy loss to predict the next visual token, and training on diverse data and model architectures, this approach demonstrates excellent scalability and performance in experimental settings.

Furthermore, Tian et al.~\cite{tian2024visual} introduced Visual Autoregressive Modeling—an innovative image generation method that improves autoregressive models by adopting a "next-scale prediction" strategy. Rather than predicting pixel-by-pixel, this approach generates images progressively from low to high resolution. This technique not only accelerates the image generation process but also captures global structure and fine-grained details more effectively, achieving higher visual quality compared to traditional autoregressive models and several popular diffusion models.

In contrast to pretraining methods that focus on pixel-level reconstruction in the image domain (e.g., MAE), the emerging paradigm of representation learning in the embedding space offers a more semantically expressive approach for geophysical modeling. Feature-space pretraining emphasizes self-supervised learning within latent representations, enabling the capture of more abstract and structurally meaningful patterns. This concept is exemplified by Meta’s I-JEPA model~\cite{assran2023self}, which embodies Yann LeCun’s vision for more human-like intelligence. Rather than predicting pixel values, I-JEPA predicts high-level semantic representations of target blocks based on a context block, thereby enhancing the model’s generalization ability in downstream tasks. In contrast, reconstruction-based methods like MAE remain grounded in the image domain, where learning is dominated by low-level local features, limiting their effectiveness in modeling complex geophysical structures.

In geophysics, such embedding-based learning frameworks can be effectively integrated with models such as Variational Autoencoders (VAEs) or seismic foundation models to establish self-supervised training pipelines for seismic images and volumetric data. Specifically, VAEs are capable of learning probabilistic embedding distributions of seismic inputs, and by optimizing the identifiability of reconstructed structures—such as geological boundaries and fault systems—they guide the model toward a geologically coherent representational space. This approach shifts the learning focus away from redundant pixel-level details and toward the capture of semantic structural features, which is particularly advantageous in dealing with seismic data characterized by strong noise and structural ambiguity. By incorporating the contextual prediction framework of I-JEPA, future work may explore conditioning on local subvolumes to predict semantic embeddings of neighboring regions, thereby improving the model’s ability to represent and generalize complex subsurface structures without relying on labeled data.

On the other hand, contrastive learning-based visual pretraining enables models to learn discriminative image representations by constructing positive and negative sample pairs. For example, in SimCLR~\cite{chen2020simple}, a positive pair is generated by applying data augmentations (such as cropping or flipping) to the same image, while other images in the batch serve as negative examples. The model learns image representations by minimizing the feature distance between positive pairs and maximizing the distance between negative pairs.

DINO V2~\cite{oquab2023dinov2} further advances visual representation learning through self-supervised contrastive learning. It employs a self-distillation mechanism based on a student-teacher architecture, where the teacher model is updated using the exponential moving average (EMA) of the student model’s parameters. Without relying on labeled data, DINO V2 learns effective feature representations by contrasting different augmented views of the same image and enforcing consistency between their representations. This approach enables the extraction of semantically meaningful and discriminative features.

Contrastive learning methods can be trained on large-scale unlabeled image datasets, offering excellent scalability. The resulting representations are well-suited for downstream tasks such as image retrieval, similarity matching, and content-based recommendation. In image retrieval, for instance, a contrastively pretrained model can extract feature embeddings from a query image and efficiently search a database for visually similar images, returning relevant results with high precision.

In geophysics, self-supervised visual model pretraining begins with the collection of large-scale visual datasets relevant to the field, such as seismic imaging, satellite imagery, remote sensing data, and subsea exploration images. These datasets contain rich spatial information and physical phenomena, providing a robust foundation for visual representation learning. During pretraining, various self-supervised learning strategies can be applied, including image reconstruction, contrastive learning, and generative modeling tasks. Image reconstruction tasks typically involve masking or distorting parts of the input image and training the model to recover the original, thereby enabling it to learn spatial structures and patterns inherent in geophysical imagery. Contrastive learning, on the other hand, teaches the model to identify similarities and differences between augmented views of the same image by comparing them across multiple perspectives or transformations.

Generative approaches—such as autoencoders or Generative Adversarial Networks (GANs)—can also be employed to model the latent structures of geophysical images, helping the model capture high-level features embedded in the data. Choosing appropriate network architectures, such as Convolutional Neural Networks (CNNs) or Vision Transformers (ViTs), enables effective extraction of both local and global features from spatial data. During training, the pretraining phase allows the model to learn general visual-semantic and physical representations from geophysical data. This is followed by a fine-tuning phase, where the model is further optimized for domain-specific tasks such as seismic wave analysis or subsurface structure identification.

In evaluating model performance, it is important not only to assess accuracy on specific tasks but also to consider the model’s generalization ability and interpretability—especially when dealing with complex geophysical imagery. These aspects ensure that the model can provide reliable insights in real-world geophysical applications. Finally, to maintain model effectiveness over time, it is essential to regularly update the training corpus, apply incremental learning, and perform task-specific fine-tuning. This ensures that the model remains aligned with ongoing advancements in geophysical research, delivering high-precision and efficient visual analysis capabilities.

\subsubsection{Multimodal Pretraining Approaches}
Multimodal pretraining involves the joint use of image and text data to train models that learn the semantic relationships between different modalities. These models are designed to understand the correspondence between visual and textual information through tasks such as image captioning (generating descriptive text for an image) and text-image retrieval (retrieving relevant images based on textual input or vice versa).

For instance, CLIP (Contrastive Language-Image Pretraining)~\cite{radford2021learning} maps images and texts into a shared embedding space using separate image and text encoders. It then employs contrastive learning to align matched image-text pairs and push apart mismatched pairs. This enables the model to perform cross-modal retrieval and understanding without relying on task-specific supervision. BLIP (Bootstrapping Language-Image Pretraining)~\cite{li2022blip} advances multimodal learning through three core objectives: vision-language alignment, image captioning, and image-text matching. By integrating these tasks, BLIP enables the model to effectively fuse visual and textual information, endowing it with both cross-modal comprehension and generation capabilities.

These multimodal pretraining approaches are highly applicable to a wide range of downstream tasks, such as visual question answering (VQA), image-to-text generation, and multimodal retrieval. In VQA, for example, the model generates accurate answers to natural language questions based on the content of an input image, enhancing the quality of human-computer interaction. Multimodal foundation models have demonstrated broad utility in domains such as social media content understanding, intelligent advertising recommendation, and digital content generation, where rich interaction between visual and textual data is essential.

On the other hand, audio-visual joint pretraining leverages both audio and visual data to train models capable of understanding the correlation and synchronization between the two modalities. For instance, in video data, the model can be trained to perform tasks such as Audio-Visual Event Localization, where it predicts the video segment corresponding to a specific audio event, or Video-to-Audio Synthesis, where it generates audio descriptions based on visual inputs.

This type of multimodal pretraining takes full advantage of the complementary nature of audio and visual information, significantly enhancing model performance in tasks such as video understanding and multimedia content analysis. It offers distinct advantages for handling complex multimedia data that contain both sound and imagery. Once trained, such models can be deployed in various real-world scenarios—for example, anomaly detection in video surveillance, where unusual events are identified using both visual and auditory cues, or video content generation, such as automatically generating suitable soundtracks for silent videos. These applications demonstrate the broad utility and potential of audio-visual multimodal models in advancing intelligent perception and content synthesis across domains.

In the field of geophysics, training multimodal self-supervised models requires the integration of diverse information sources, such as seismic waveforms, remote sensing imagery, subsurface electromagnetic data, and meteorological measurements. By designing appropriate multimodal learning tasks, models can learn the intrinsic relationships and complementary nature of different data modalities. Multimodal data fusion techniques enable the unified representation of heterogeneous inputs—such as images, text, and time-series data—within a shared embedding space. This allows the model to jointly process and reason over varied geophysical data types. For instance, contrastive learning between image-text pairs can establish semantic associations between seismic images and their corresponding textual descriptions, allowing the model to understand the connection between physical phenomena and their linguistic representations.

Self-supervised learning tasks, such as image-text matching, data reconstruction, or multimodal contrastive learning, facilitate the automatic extraction of features from large-scale unlabeled geophysical datasets. Through such tasks, the model can capture spatial patterns, temporal dynamics, and underlying physical relationships inherent to geophysical processes. Furthermore, multimodal models can employ unified encoding architectures—such as multimodal Transformers—to process different modalities simultaneously. By leveraging a shared representation space, these models enhance cross-modal understanding and inference capabilities.

After pretraining, the model can be fine-tuned for specific downstream geophysical applications, such as subsurface resource exploration, seismic wave prediction, or disaster early warning systems. With continual training and data updates, the model can adapt to the evolving characteristics of geophysical data and task requirements. Ultimately, this approach enables robust multimodal data analysis and prediction, offering deeper insights and more accurate decision support for geophysical research and practice.

Despite their potential, the application of multimodal pretraining methods in geophysics faces numerous challenges. These primarily involve issues related to data fusion, cross-modal relationship modeling, data annotation and quality, model scalability and computational complexity, and generalization performance. Geophysical data are characterized by high diversity in source and format, making it difficult to effectively fuse different modalities—such as seismic data, remote sensing imagery, and electromagnetic measurements—while maintaining information consistency and maximizing model performance. Achieving seamless integration of such heterogeneous data remains a central challenge. Additionally, the high cost and subjectivity associated with data annotation often lead to inconsistencies in training data quality. Effective modeling of cross-modal relationships and long-range dependencies still requires further methodological advances to improve model robustness and representation capacity. The significant computational demands of training large-scale multimodal models on massive geophysical datasets also raise concerns about efficiency and scalability. Optimizing training strategies and model architectures to reduce resource consumption without compromising performance is thus a critical area of research. Lastly, integrating domain-specific physical knowledge into deep learning frameworks remains a key open challenge. Ensuring that model outputs conform to known physical laws and constraints is essential for reliability and scientific interpretability, particularly in high-stakes geophysical applications.

\section{Post-Training and Deployment of Foundation Models}
\label{sec:applications}

\begin{figure*}[!htb]%
\centering
\includegraphics[width=1\textwidth]{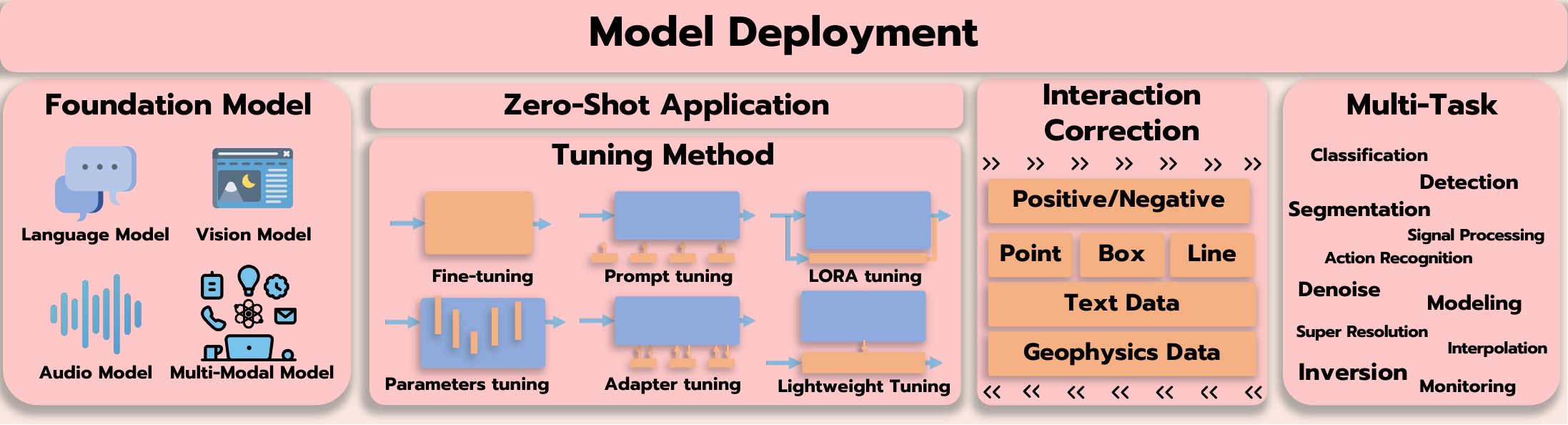}
\caption{Deployment workflow of geophysical foundation models. This figure illustrates deployment strategies for foundation models across language, vision, audio, and multimodal tasks. It outlines several approaches, including direct application and various fine-tuning techniques—such as full fine-tuning, prompt tuning, parameter-efficient tuning, adapter tuning, and lightweight tuning—as well as interactive refinement methods. The workflow culminates in applications to multi-task scenarios, including classification, detection, segmentation, signal processing, modeling, and inversion. The figure highlights the broad applicability and flexibility of foundation models in the geophysical domain.
}
\label{fig:ModelDeployment}
\end{figure*}

Following the in-depth discussion of foundation model pretraining, we now focus on the post-training and deployment phase. This stage encompasses not only task-specific fine-tuning but also performance optimization to ensure scalability, efficiency, and practical usability of the model in real-world applications. After pretraining and initial fine-tuning for specific tasks, the core objective of post-training is to further adapt the model to handle the complexity and heterogeneity of geophysical data effectively. By employing techniques such as full fine-tuning, prompt tuning, and adapter tuning, the model can achieve improved performance on specialized geophysical tasks. In practice, post-training requires carefully tailored adjustments that reflect the unique characteristics of geophysical datasets, along with a focus on interactive optimization—leveraging multi-model collaboration to enhance task execution efficiency. Moreover, as the scale of foundation models continues to grow, achieving lightweight and flexible deployment becomes essential for efficient application. Techniques such as model distillation, pruning, and quantization can reduce computational and storage overhead while maintaining high performance. These optimizations facilitate scalable deployment across tasks and regions, ensuring that foundation models can be effectively applied in diverse geophysical scenarios (Fig.~\ref{fig:ModelDeployment}).

\subsection{Fine-Tuning Methods}
In the application of industry-scale foundation models, the combination of pretraining and fine-tuning strategies is essential to ensure effective performance across diverse tasks. While large models pretrained on massive datasets exhibit strong generalization capabilities, achieving optimal results in specific domains or tasks requires further refinement through appropriate fine-tuning techniques.

Fine-tuning is the most commonly used approach, wherein all parameters of the pretrained model are updated during continued training to adapt to the new task. When sufficient labeled data is available, this method enables comprehensive adjustment of the model’s weights, thereby maximizing task-specific performance. However, fine-tuning is computationally intensive—particularly for large-scale models—and demands substantial hardware resources. Additionally, in low-data regimes, it may lead to overfitting, limiting its generalization ability on unseen examples.

In the fine-tuning of pretrained models, several mainstream approaches have emerged, including Prompt Tuning\cite{brown2020language}, Adapter Tuning\cite{houlsby2019parameter}, LoRA (Low-Rank Adaptation)\cite{hu2022lora}, Prefix Tuning\cite{li2021prefix}, and P-Tuning~\cite{liu2021p}, each suited to specific scenarios and application domains. Prompt Tuning modifies only the input prompts rather than the model parameters. It is particularly well-suited for zero-shot and few-shot tasks, offering simplicity and strong adaptability. However, its performance may be limited on more complex tasks that require deeper model adjustments. Adapter Tuning introduces small trainable modules between layers of the pretrained model, enabling the tuning of a limited number of parameters. It is especially suitable for multi-task scenarios where rapid deployment and low computational cost are priorities. Nonetheless, adapter-based approaches may underperform compared to full fine-tuning in highly complex tasks. LoRA reduces the number of trainable parameters by applying low-rank decomposition to weight updates. It is ideal for scenarios with limited computational resources but high performance requirements, offering substantial parameter efficiency. However, it may lack flexibility in some use cases compared to other techniques. Prefix Tuning prepends learnable prefix vectors to the input sequence, making it suitable for multi-task settings where a shared backbone model is desirable. This approach reduces both storage and computational overhead, although its task-specific performance can be inferior to more expressive fine-tuning methods. P-Tuning and P-Tuning V2 enhance task performance by injecting continuous learnable vectors into the model’s input embedding space. While P-Tuning is primarily designed for unimodal tasks, P-Tuning V2 extends the approach to multimodal settings and demonstrates stronger performance on complex tasks. However, these methods still face scalability challenges when applied to very large models or datasets. In the following section, we will explore how these fine-tuning strategies can be effectively adapted for geophysical data and tasks.

\subsubsection{Fine-Tuning}

Fine-tuning is a conventional and widely adopted approach in model adaptation, in which all—or a substantial portion—of the pretrained model's parameters are updated to better align with the requirements of a specific downstream task. Unlike lightweight fine-tuning methods, full fine-tuning directly modifies the model's internal weights, enabling optimal task-specific performance. The core principle of fine-tuning lies in adjusting the pretrained model to accommodate the distribution of new data, thereby significantly enhancing its effectiveness on the target task. This method is particularly suited to scenarios where abundant labeled data are available, allowing for deep, task-specific customization. As a result, full fine-tuning tends to excel in complex tasks that demand high model expressiveness and adaptability.

One of the key advantages of fine-tuning lies in its strong adaptability to single tasks. By thoroughly updating all model parameters, fine-tuning enables highly granular optimization, allowing the model to precisely capture the patterns and features relevant to a target task. However, this level of customization comes with significant computational and storage costs—especially when fine-tuning large-scale models—often requiring substantial memory and processing resources. As such, fine-tuning is better suited for high-resource environments and critical tasks that demand dedicated optimization. Fine-tuning has become an indispensable method in many machine learning and natural language processing applications, including sentiment analysis, question answering, and text classification. Despite its computational demands, it offers unparalleled customization capability, making it the preferred strategy for complex, domain-specific tasks. It provides a powerful solution for adapting pretrained foundation models to specific use cases through deep optimization.

In the context of geophysical foundation models, fine-tuning with domain-specific data is a crucial step for enhancing model performance. First and foremost, data preparation and preprocessing are essential. This includes collecting high-quality, labeled geophysical datasets and applying normalization or standardization to ensure that the inputs conform to consistent scales and distributions. In addition, data augmentation techniques—such as noise injection, translation, or rotation—can be employed to improve generalization, particularly when labeled data are scarce.

In selecting a fine-tuning strategy, one can choose between full-model fine-tuning and partial fine-tuning, depending on the characteristics of the pretrained model and the requirements of the downstream task. If the foundation model has been pretrained comprehensively and the downstream task closely resembles the pretraining objectives, full-model fine-tuning can be applied directly. In this case, all model parameters are updated via backpropagation, allowing the model to better adapt to geophysical data.

However, when the amount of training data is limited or the downstream task differs significantly from the pretraining task, partial fine-tuning is often recommended. This approach involves freezing the earlier layers of the model and updating only the later layers, thereby reducing computational cost and mitigating the risk of overfitting. During the fine-tuning process, both supervised learning and self-supervised learning are commonly employed. In cases where ample labeled geophysical data are available, supervised fine-tuning can be performed by minimizing the loss between model predictions and ground-truth labels, gradually optimizing model performance. When labeled data are scarce, self-supervised learning serves as an effective alternative. Using generative objectives or contrastive learning, the model can extract meaningful representations from unlabeled geophysical data, enhancing its ability to generalize to downstream tasks.

In addition, multi-task learning can further enhance model performance. By simultaneously addressing multiple tasks—such as seismic signal inversion and subsurface property prediction—within a single model, multi-task learning enables the extraction of richer geophysical features. This approach not only improves performance on individual tasks but also increases the model's generalization capability across diverse geophysical scenarios.

During fine-tuning, using a smaller learning rate is an important strategy to prevent overfitting. Since pretrained models already possess strong feature extraction capabilities, a high learning rate may lead to catastrophic forgetting of previously acquired knowledge. To further improve generalization, especially in low-data settings, regularization techniques such as L2 regularization and dropout are commonly applied. These methods help constrain the model's complexity and reduce the risk of overfitting.

After fine-tuning, the model's performance should be evaluated using a dedicated validation set. Depending on the task, evaluation metrics such as accuracy, recall, and F1-score can be used to provide a comprehensive assessment. Hyperparameters may be adjusted based on validation results to optimize performance. Techniques such as cross-validation can also be employed to ensure the robustness of the model across different data subsets. Once optimization is complete, ensemble learning methods can be utilized to further boost model performance by combining predictions from multiple fine-tuned models. Ultimately, the optimized and validated models can be deployed for real-world geophysical applications, such as seismic inversion, mineral exploration, and other subsurface imaging tasks.

\subsubsection{Efficient Fine-Tuning Methods}

In geophysical data processing, efficient fine-tuning strategies are essential due to the complexity, heterogeneity, and scarcity of geophysical data. These datasets span a wide range of tasks—from hydrocarbon exploration to seismic velocity prediction—and often involve limited annotations and constrained computational resources. To enable fast and accurate model adaptation under these conditions, it is crucial to select fine-tuning methods that are well aligned with the characteristics of the data and the demands of the task~\cite{liu2024visual}. For instance, Prompt Tuning\cite{brown2020language} is well-suited to low-resource scenarios, where task-specific prompts guide the model to adapt rapidly to new tasks with minimal labeled data. Adapter Tuning\cite{houlsby2019parameter}, which inserts small trainable modules into a frozen pretrained model, is highly effective for multi-task learning and can enhance adaptability across diverse geophysical data modalities. LoRA (Low-Rank Adaptation)~\cite{hu2022lora} reduces the number of trainable parameters via low-rank matrix decomposition, making it ideal for large-scale data processing while reducing computational overhead. Given the high dimensionality and noise characteristics of geophysical data, effective preprocessing and feature selection are vital prior to fine-tuning. Techniques such as Prefix Tuning\cite{li2021prefix} and P-Tuning\cite{liu2021p} offer robust mechanisms for integrating multimodal data, improving the model's ability to generalize across different geophysical data sources. Adapter-based approaches also support cross-task and cross-domain transfer, further enhancing generalization.

Integrating physics-based constraints and domain knowledge—such as wave equations or seismic wave propagation models—into the fine-tuning process can significantly improve predictive accuracy and model interpretability. Since geophysical applications often involve large-scale datasets, combining LoRA tuning with distributed training techniques can reduce memory usage and computational costs, leading to more efficient fine-tuning pipelines. Finally, continuous fine-tuning and adaptive updates, as implemented in methods like P-Tuning v2, are particularly advantageous in long-term geophysical monitoring projects. These techniques allow the model to accommodate evolving geological features and exploration targets over time, maintaining robustness and adaptability. In summary, the choice of an efficient fine-tuning strategy should be based on a comprehensive consideration of task objectives, available computational resources, and the intrinsic properties of the geophysical data. Such flexibility ensures optimal model performance under real-world constraints.

During the fine-tuning stage of geophysical models, a range of unique requirements and challenges must be addressed—stemming primarily from the complexity of the data, the diversity of tasks, and the constraints of computational resources. First, geophysical data often suffer from a lack of large-scale labeled datasets, particularly in tasks such as deepwater hydrocarbon exploration and subsurface structure prediction. Labeling these data is expensive and requires domain expertise, making it especially challenging to enhance model performance in low-resource scenarios, including zero-shot and few-shot settings. Second, geophysical data originate from a variety of sensors and acquisition methods, leading to significant heterogeneity—for example, among seismic, gravity, and remote sensing datasets. Effectively fusing such multimodal data and enabling multimodal applications is a key challenge in the fine-tuning process. In addition, geophysical data are typically high-dimensional and noisy, especially in the case of 3D seismic volumes and high-resolution remote sensing imagery. Denoising, feature selection, and computational optimization are thus critical to improving both efficiency and accuracy.

Another major challenge arises from the fact that geophysical data are governed by strict physical laws—such as wave equations and models of seismic wave propagation. Incorporating these physical constraints into the fine-tuning process is essential to ensure that model outputs are physically meaningful and scientifically reliable. Furthermore, geophysical tasks are often cross-domain and cross-task in nature. For example, hydrocarbon exploration and seismic velocity prediction may be related but differ significantly in their objectives and data characteristics. Designing fine-tuning methods that support transfer learning while avoiding catastrophic forgetting is essential for robust performance across tasks. Compounding these issues is the high computational cost often associated with processing geophysical datasets. Efficiently fine-tuning large-scale pretrained models while reducing computational and storage overhead—especially in resource-constrained environments—remains a fundamental technical bottleneck. Lastly, geophysical data are subject to continuous updates over time. Maintaining model adaptability through continual fine-tuning and self-adaptive updates is crucial to ensuring long-term performance stability as new geological formations or exploration targets are introduced.

In summary, the fine-tuning phase in geophysical applications must contend with challenges including data scarcity, multimodal fusion, physical consistency, task heterogeneity, and resource limitations. Therefore, fine-tuning strategies must be both flexible and efficient, while also integrating domain knowledge and physical priors to ensure optimal performance under real-world constraints.

\subsubsection{Geological and Geophysical Prior-Guided Fine-Tuning}

In the fine-tuning of foundation models, the effective incorporation of geological and geophysical prior knowledge is a critical step to ensure that the model not only performs well in a data-driven sense but also produces outputs consistent with established physical and geological principles. While foundation models, pretrained on large-scale datasets, are capable of capturing latent patterns in the data, purely data-driven approaches are often insufficient to guarantee physical plausibility in domain-specific geophysical applications. Therefore, integrating geological and geophysical priors during fine-tuning becomes essential for enhancing the model’s domain specialization, interpretability, and predictive accuracy.

\begin{figure*}[!htb]%
\centering
\includegraphics[width=1\textwidth]{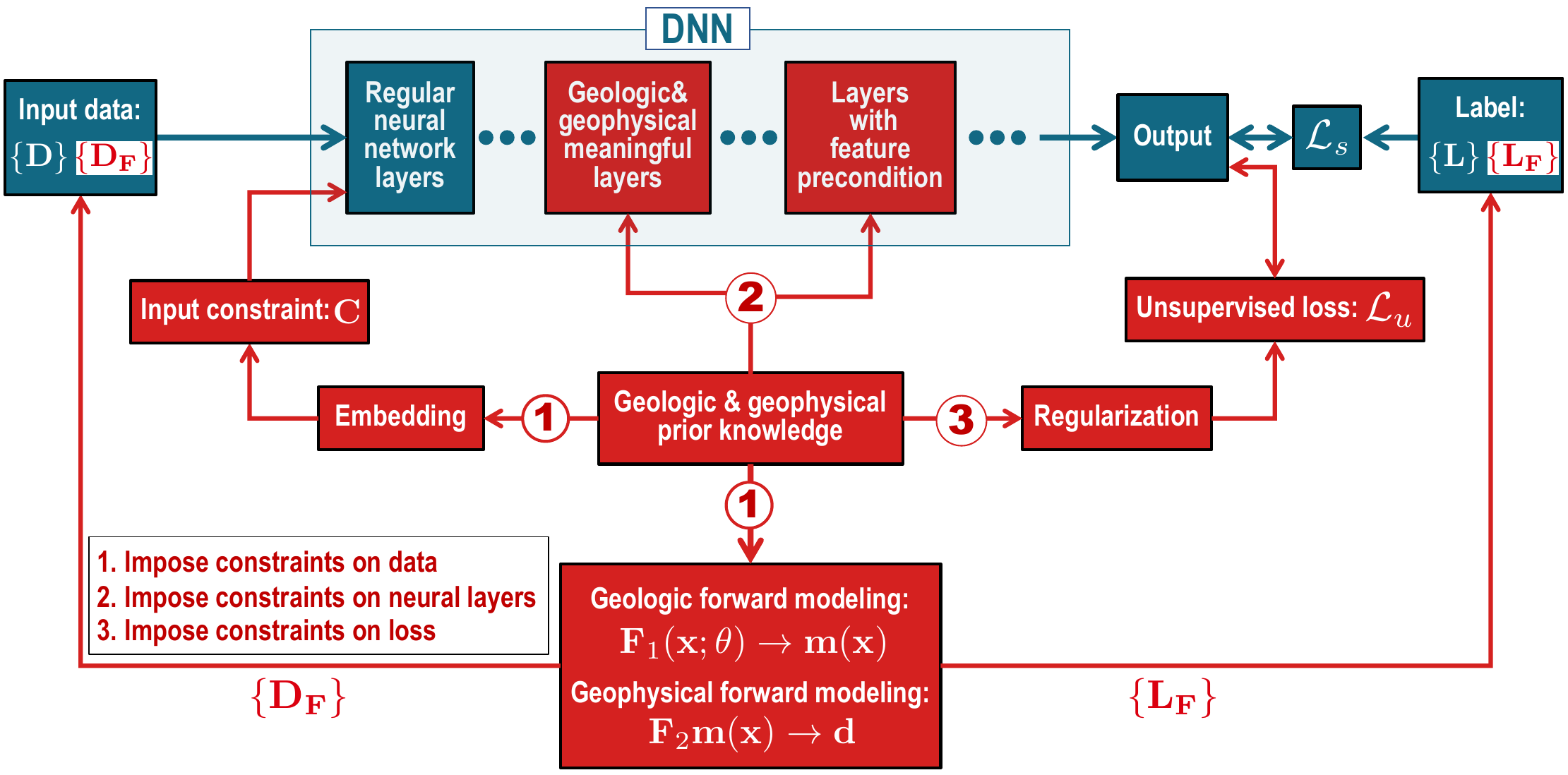}
\caption{\citep{wu2023sensing} introduced three representative strategies for integrating geological and geophysical prior knowledge (marked in red) into deep neural network (DNN) architectures (in blue). (1) Data-level integration involves embedding expert knowledge into input features or generating synthetic datasets via forward modeling informed by geological and geophysical principles. (2) Network-level integration modifies the model structure itself, for example by introducing geophysically meaningful layers or conditioning intermediate feature representations. (3) Loss-level integration incorporates prior constraints into the optimization objective using customized loss functions.
}
\label{fig: Prior}
\end{figure*}

In practical applications, it is essential to construct task-specific loss functions tailored to the geophysical domain (Figure \ref{fig: Prior}). These loss functions should not only retain the optimization characteristics of traditional data-driven objectives but also explicitly incorporate geological and geophysical priors, including relevant physical constraints. Conventional loss functions—such as Mean Squared Error (MSE)—often fall short when modeling complex geological structures and governing physical laws. By integrating known geological horizons, lithological distributions, and wave propagation properties into the loss function as regularization terms or auxiliary constraints, models can be guided to produce physically consistent and geologically plausible predictions. For example, in seismic velocity inversion tasks, incorporating known stratigraphic boundaries or lithological information as physical constraints within the loss function can improve the credibility of the inversion results. Moreover, physics-informed neural networks (PINNs) offer a compelling framework for embedding physical laws directly into the training objective. By including residuals of governing equations—such as the wave equation or Maxwell’s equations—as part of the loss function, PINNs ensure that model optimization is driven not only by data fidelity but also by adherence to fundamental geophysical mechanisms. In seismic data analysis, for instance, embedding the residuals of the acoustic or elastic wave equation into the loss function significantly enhances the stability and interpretability of the model outputs, while also reducing the likelihood of physically implausible predictions often encountered in purely data-driven approaches. Furthermore, given the inherent multi-scale nature of geophysical data, the design of the loss function should adopt a multi-scale optimization strategy. Specifically, the model should be trained to simultaneously capture both low-frequency and high-frequency components of the data. This enables the model to represent the large-scale geological structures as well as fine-grained local features. By balancing global geological trends with detailed local variations, multi-scale loss functions help prevent overly smoothed predictions and reduce the risk of neglecting critical structural information.

The heterogeneity of geophysical data sources—such as seismometers, gravimeters, and magnetometers—leads to discrepancies in physical scales and data characteristics. These differences can introduce inconsistencies in model predictions. To address this, it is essential to include data consistency loss terms in the objective function. These terms leverage prior geological models or known lithological distributions to enforce physical consistency across modalities and sensor types, ensuring that outputs from different data sources are aligned with one another. Moreover, due to the spatiotemporal variability of geological and geophysical data, incorporating spatiotemporal normalization terms into the loss function during fine-tuning can further enhance training stability and model generalization. These terms help reduce inconsistencies in data resolution, sampling rate, or spatial extent, allowing the model to learn more coherent and robust representations across different temporal and spatial scales.

To enable effective knowledge sharing across different geophysical tasks, multi-task learning strategies can be incorporated during the fine-tuning process. For example, by training a unified network architecture to perform inversion tasks across multiple geophysical data types—such as seismic, gravity, and magnetic field data—the model can promote the transfer and integration of geological and physical information among modalities. By designing task-specific physical constraints or prior knowledge for each modality and integrating them into a joint loss function, the model’s overall predictive accuracy and generalization performance can be significantly improved.

In fine-tuning foundation models for geophysical applications, systematically incorporating geological and geophysical priors—through the construction of loss functions that integrate physical constraints, multi-scale optimization, data consistency, and multi-task learning—is a powerful strategy for improving both accuracy and physical plausibility. This comprehensive approach enhances the model’s robustness and interpretability in geophysical contexts, ensuring that predictions are not only data-driven but also grounded in domain-specific physical principles.

Ultimately, such fine-tuning strategies increase the generalization capability and practical value of foundation models, enabling reliable deployment in real-world geophysical tasks with complex data and limited supervision.

\subsection{Model Interaction}
Model interaction is a technique that enhances overall model performance through collaboration and information sharing among multiple models, and it is particularly well-suited for complex and multimodal task environments. The core idea is to enable deep coordination across models, allowing them to exchange features, strategies, or intermediate representations based on the specific demands of the task. This cooperative paradigm often leads to superior performance compared to the isolated operation of individual models. For example, architectures inspired by the Segment Anything Model (SAM) demonstrate how model interaction can achieve high-precision results in image segmentation. Beyond segmentation, such architectures can also interact with other models—such as language models or object detection models—to extend the system’s adaptability. This integration supports cross-modal interaction and enables significant performance improvements in tasks requiring multimodal reasoning and understanding.

Through model interaction, systems can dynamically leverage the specialized capabilities of individual models, establishing a form of “collaborative specialization”. For example, in image segmentation tasks, Segment Anything Model (SAM) can receive textual prompts or contextual semantics from a language model, enabling more refined segmentation with enhanced semantic consistency. In cross-modal scenarios, SAM can also interact with object detection models by integrating information about detected objects within the image, thereby achieving significant improvements in fine-grained object segmentation. Such interactive architectures create information bridges between models, allowing flexible transmission and integration of features across different components. This collaborative framework enhances the system’s ability to adapt to more complex and heterogeneous task requirements, supporting richer multimodal reasoning and more precise output generation.

In the development of geophysical models, incorporating interaction between models and expert feedback is a critical component for improving model performance and reliability. Expert input allows model outputs to be iteratively refined and adjusted, ensuring that predictions not only comply with physical laws but are also better aligned with practical application scenarios. For example, in seismic data inversion, geophysicists can evaluate discrepancies between the model-predicted subsurface structures and actual observational data, offering domain-specific insights that are difficult for the model to infer autonomously. The model, in turn, can integrate this feedback to adjust its parameters or inference strategy, thereby optimizing the prediction results.

This feedback loop enables the model to dynamically adapt during deployment, improving its accuracy and robustness in complex geophysical environments. Moreover, the interactive process facilitates the gradual infusion of expert knowledge into the model’s training and reasoning stages, contributing to the evolution of a continually improving intelligent system. Such expert-model interaction not only enhances the quality and trustworthiness of the outputs but also bridges the gap between data-driven algorithms and domain-driven geoscientific understanding.

Interactive development of geophysical models is a multi-level, multi-faceted process involving the integration of expert knowledge, machine learning, and data analysis, with the goal of continuously optimizing model performance through feedback from domain experts. A critical first step in this process is the translation of geophysicists’ expertise into model-compatible forms—such as rule-based systems, empirical heuristics, or physical constraints—so that the model can dynamically incorporate expert feedback and adjust its behavior accordingly.

Through expert corrections, the model can better align with real-world scenarios. For example, in seismic inversion or subsurface reservoir prediction, geophysicists can refine model outputs by identifying discrepancies based on observed anomalies or geological context. The core of this interactive process lies in a real-time feedback mechanism, in which experts provide both qualitative and quantitative evaluations of model outputs. Qualitative feedback offers intuitive interpretations based on expert understanding, while quantitative feedback proposes data-driven corrections grounded in physical measurements or prior knowledge.

This feedback loop enables rapid, targeted refinement of the model, leading to improved accuracy and practical relevance. With the rapid advancement of artificial intelligence and machine learning, the integration of expert systems and learning algorithms has become a powerful approach to model enhancement. While machine learning excels at uncovering patterns from large datasets, expert systems inject domain-specific priors that prevent the model from converging to suboptimal solutions. Furthermore, machine learning techniques can serve as rapid diagnostic tools, assisting experts in identifying latent issues and guiding the refinement of both data and model structure.

Moreover, interactive geophysical modeling places a strong emphasis on data and model result visualization. Tools such as 3D visualization allow experts to clearly observe the model's reasoning process and local prediction outcomes, thereby enabling more effective and targeted feedback. Geophysical data often involve multimodal information, such as seismic waveforms and geomagnetic measurements. Effective fusion of these heterogeneous data types requires models capable of integrating inputs from diverse sources and adapting their parameters and strategies based on expert feedback—ultimately enhancing the accuracy of subsurface imaging.

To support continuous improvement, some geophysical models incorporate self-learning capabilities, allowing them to accumulate and internalize feedback through iterative expert interaction. Over time, the model refines itself and adapts to the evolving geophysical environment. To fully realize this potential, automation and integration must be embedded into the interactive development workflow. Expert feedback should be seamlessly incorporated into the model pipeline, enabling rapid iterations and efficient optimization.

In large-scale geophysical applications, such automation not only conserves expert time but also significantly boosts overall productivity. Through this iterative and interactive development process, geophysical models become progressively more accurate, interpretable, and robust—ultimately evolving into reliable decision support tools for exploration, monitoring, and risk assessment.

\subsection{Instruction Learning and Reinforcement Learning}

\begin{figure*}[!htb]%
\centering
\includegraphics[width=1\textwidth]{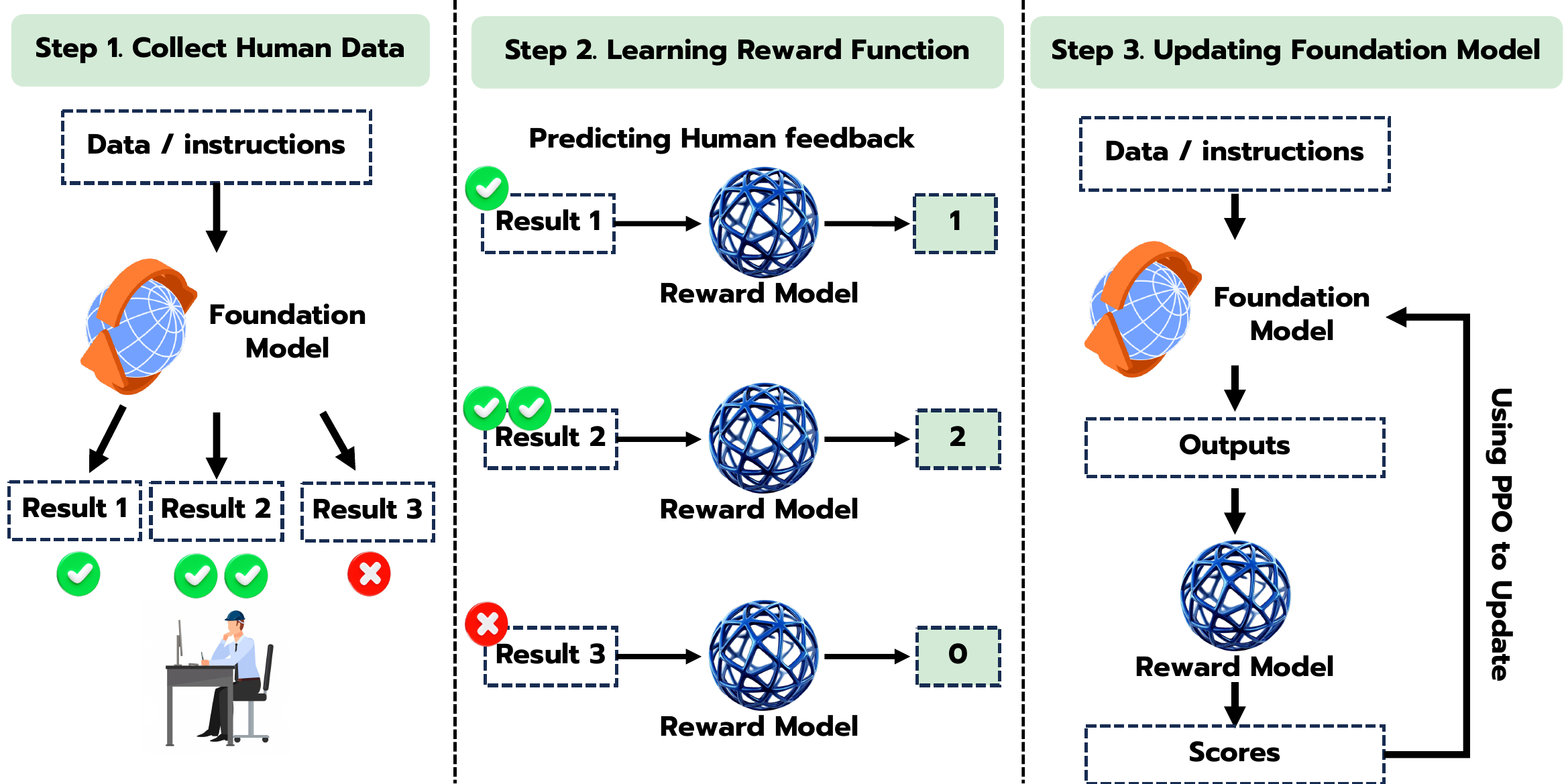}
\caption{Instruction Tuning and Reinforcement Learning for Geophysical Foundation Models.
This figure illustrates the workflow of reinforcement learning with human feedback (RLHF) applied to instruction tuning in geophysical foundation models. First, human-annotated geophysical instruction data is collected (Step 1). Next, a reward model is trained on this data to predict human preferences for model outputs (Step 2). Finally, based on the reward model’s evaluations, the geophysical foundation model is updated and optimized using the Proximal Policy Optimization (PPO) algorithm (Step 3).
}
\label{fig:Instruct}
\end{figure*}

In the field of geophysics, the integration of instruction learning and reinforcement learning offers promising opportunities for innovative applications—particularly in solving complex problems and addressing data-scarce scenarios. Instruction learning can provide explicit procedural guidance for geophysical inversion, enabling models to follow a structured workflow when processing various types of geophysical data and incrementally refining subsurface models (Fig.~\ref{fig:Instruct}).

Reinforcement learning, on the other hand, introduces a reward-driven mechanism that encourages the model to explore diverse inversion strategies and make optimal decisions at each stage of the process. This can help reduce bias and uncertainty commonly encountered during inversion, especially in ill-posed or noisy settings. The combination of these two paradigms not only enhances inversion accuracy but also reduces the need for manual intervention, leading to a more efficient and adaptive inversion pipeline. This hybrid approach holds significant potential for improving performance in complex geophysical environments while increasing model autonomy and robustness.

In the context of seismic activity forecasting and disaster early warning, instruction learning can guide the model to systematically extract meaningful features from high-dimensional seismic data and perform step-by-step prediction. Building on this foundation, reinforcement learning introduces a self-optimization mechanism that adaptively refines prediction strategies, thereby improving the accuracy and stability of the model. By learning from both historical and real-time data, reinforcement learning allows the model to incrementally identify the most effective prediction pathways, optimizing both the timing and precision of early warnings. As the model interacts with the evolving data environment, it can adjust its predictive behavior to reduce false alarms and missed detections, thus offering intelligent support for disaster warning systems. In subsurface resource exploration, the integration of instruction learning and reinforcement learning also proves highly beneficial. Instruction learning can serve as a procedural guide, ensuring that each step in the exploration process follows an optimal sequence. Reinforcement learning, meanwhile, continuously adapts to the exploration environment by adjusting data acquisition strategies, exploration depth, and sampling locations to maximize exploration efficiency and resource yield. Through this self-learning process, reinforcement learning empowers the model to respond effectively to challenges such as complex subsurface structures and inconsistent physical measurements. As a result, the model achieves greater accuracy and cost-effectiveness in subsurface resource exploration, even under uncertainty.

In automated processing and analysis of geophysical data, instruction learning can assist models in autonomously performing tasks such as data preprocessing and feature extraction, thereby reducing the need for manual intervention. Reinforcement learning complements this by enabling the model to optimize its processing strategies through self-feedback. It can iteratively refine data handling algorithms, sampling methods, and transformation procedures to improve the efficiency and accuracy of geophysical data analysis. For instance, in processing complex seismic datasets, reinforcement learning can evaluate the effectiveness of different algorithms and progressively optimize their application during preprocessing—thereby enhancing the model’s capability to handle large-scale data.

The integration of multimodal geophysical data also stands to benefit from the synergy of instruction learning and reinforcement learning. Instruction learning provides procedural guidance for effectively processing and integrating diverse geophysical modalities, such as seismic, gravity, and magnetic field data. Reinforcement learning then supports optimal decision-making during the fusion process, dynamically adjusting the relative weighting of different data sources. By doing so, reinforcement learning ensures that the model can adaptively synthesize multimodal inputs and generate more accurate and geologically consistent subsurface models. This approach improves the model's capacity to understand complex physical characteristics of the subsurface environment, thereby enhancing its predictive performance and robustness in integrated geophysical interpretation.

In applications such as subsurface monitoring and intelligent detection, the combination of instruction learning and reinforcement learning can significantly enhance a model’s adaptability. For instance, in groundwater or gas leak monitoring, instruction learning can guide the model through the key procedures and methodologies required for effective monitoring. Reinforcement learning then leverages real-time feedback to continuously optimize detection strategies—adjusting parameters such as monitoring frequency and depth—to achieve optimal performance.

Reinforcement learning enables the model to dynamically adapt its behavior based on changes in real-time data, improving the sensitivity and accuracy of subsurface monitoring systems, especially in complex or hard-to-observe environments. In the broader context of intelligent decision support systems in geophysics, the integration of instruction learning and reinforcement learning also plays a vital role in optimizing decision-making processes. Instruction learning helps the model extract relevant information from various data sources and generate actionable decision instructions. Meanwhile, reinforcement learning continuously updates and refines these strategies through a reward-based mechanism to achieve task-specific objectives. For example, in safety assessments of underground infrastructure, reinforcement learning can use historical data and real-time monitoring feedback to learn optimal response strategies, thereby improving emergency responsiveness and reducing potential risks.

\subsection{Lightweight and Flexible Deployment}
\begin{figure*}[!htb]%
\centering
\includegraphics[width=1\textwidth]{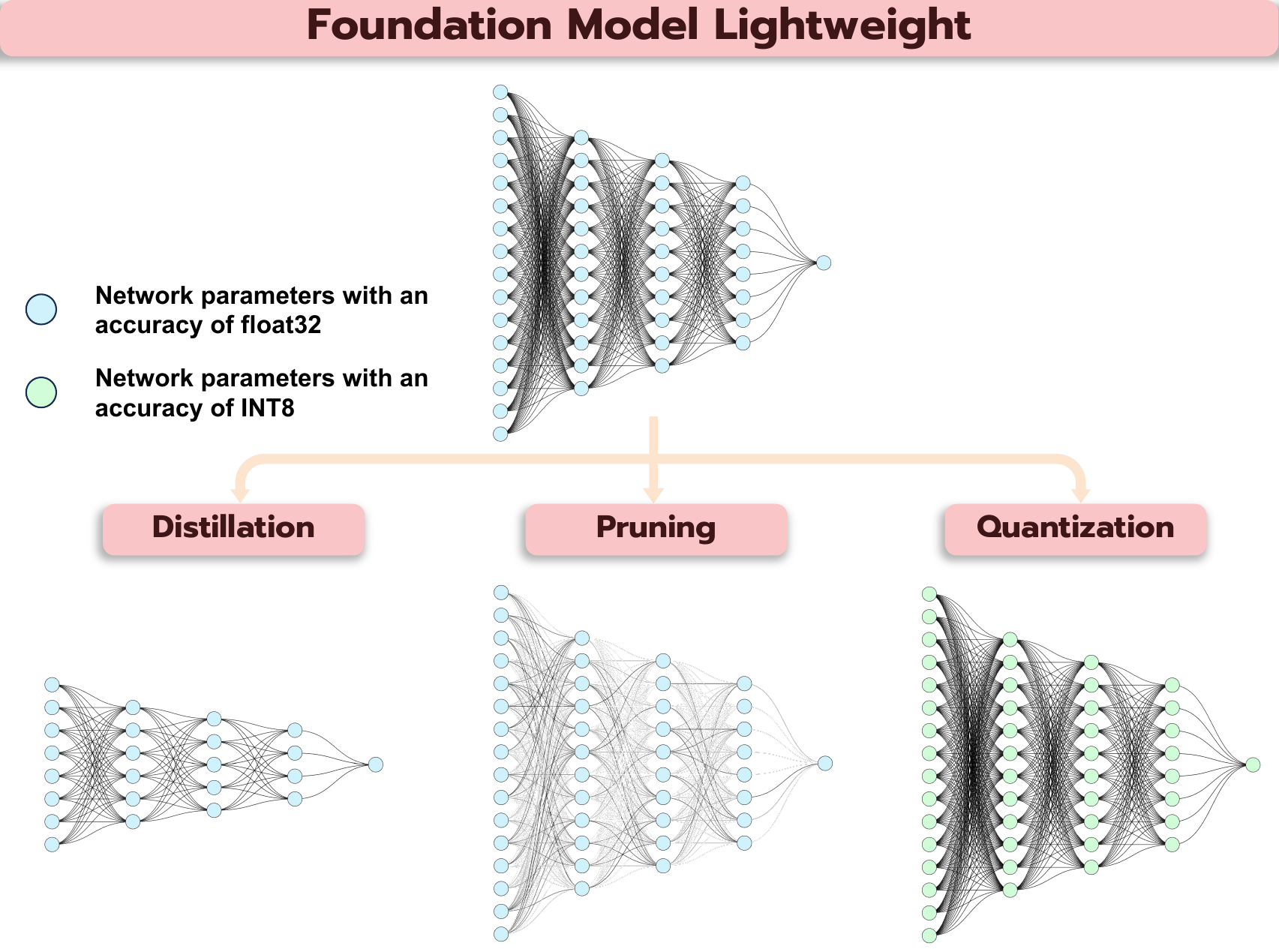}
\caption{Techniques for Lightweight Foundation Model Optimization. A full-scale foundation model (left) is streamlined using three mainstream compression strategies: Distillation, Pruning, and Quantization. Distillation transfers knowledge from the full model to a smaller one, pruning removes less critical parameters to reduce complexity, and quantization reduces numerical precision (e.g., float32 to INT8) to improve efficiency. Orange nodes represent parameters with float32 precision, while green nodes represent INT8 precision. These techniques are crucial for deploying foundation models on resource-constrained devices without significant performance degradation.
}
\label{fig:lightweight}
\end{figure*}

Foundation models typically contain tens to hundreds of billions of parameters, posing significant challenges for real-world deployment. To reduce computational overhead and improve inference speed, several model optimization techniques are often employed (Figure \ref{fig:lightweight}).

\subsubsection{Distillation}

Knowledge distillation is a model compression technique that transfers knowledge from a large pretrained model (referred to as the teacher model) to a smaller, more efficient model (known as the student model), thereby reducing computational cost and memory usage while preserving task performance (The first column in Figure \ref{fig:lightweight}). The core idea is to leverage the rich representational capacity of the teacher model to guide the training of the student model, enabling the latter to achieve comparable performance with significantly fewer parameters.

Unlike training a small model from scratch, distillation provides the student model with soft labels (i.e., probability distributions over outputs) or intermediate feature representations from the teacher model. These richer training signals help the student model learn finer-grained patterns and structural relationships within the data.

The distillation process typically involves two stages: first, the teacher model is trained on large-scale datasets to achieve strong performance; second, the student model is trained to mimic the outputs of the teacher, either at the output layer (e.g., matching prediction probabilities) or at intermediate layers (e.g., replicating hidden feature maps). Compared to direct fine-tuning or parameter-sharing approaches, distillation is especially suitable for resource-constrained environments or scenarios with strict inference latency requirements. A major advantage of distillation is its ability to preserve the knowledge of large-scale models while significantly reducing model size, thereby achieving a practical trade-off between performance and efficiency.

The distillation process typically consists of two main stages: first, the teacher model is trained using a standard optimization procedure on large-scale datasets to achieve strong performance. Next, the student model is trained to replicate the outputs of the teacher model—such as prediction probability distributions or intermediate feature representations—to learn task-specific knowledge. Compared to direct fine-tuning or parameter-sharing strategies, distillation is particularly well-suited for scenarios with limited computational resources or those requiring low-latency inference. One of the key advantages of distillation lies in its ability to substantially reduce model size while retaining much of the teacher model's knowledge, thereby achieving an effective balance between performance and efficiency.

\subsubsection{Pruning}

Pruning is a model compression and acceleration technique that reduces the computational complexity and memory footprint of a pretrained model by removing unimportant or redundant parameters, while maintaining comparable performance (The second column in Figure \ref{fig:lightweight}). The core idea behind pruning is that not all parameters or connections in a neural network contribute equally to the final output. Therefore, parameters with minimal influence on prediction can be eliminated without significantly degrading model accuracy. Similar to knowledge distillation, pruning is primarily employed to address deployment-phase optimization challenges, particularly in scenarios with constrained computational resources or strict latency requirements. By effectively reducing model size and inference cost, pruning enables more efficient and scalable deployment of large-scale models, making it especially valuable for geophysical applications requiring real-time or resource-aware processing.

The implementation of pruning encompasses various strategies, including weight pruning, structural pruning, and layer-wise pruning. Weight pruning removes individual weights in the parameter matrices that are close to zero, thereby reducing computational overhead at a fine-grained level. Structural pruning takes a more systematic approach by eliminating entire neurons, channels, or filters, leading to a more compact and interpretable network architecture. Layer-wise pruning simplifies the network at the architectural level by reducing the depth or width of the model, which directly accelerates inference. These strategies can be applied independently or in combination to maximize model compression while maintaining acceptable performance on downstream tasks.

In deep learning, pruning is widely used to remove network components with minimal impact on output, thereby improving efficiency in resource-constrained environments. For example, in convolutional neural networks (CNNs), pruning can eliminate redundant convolutional filters or channels, enabling deployment on embedded systems with limited memory and compute. In natural language processing (NLP), pruning helps reduce the parameter count of large pretrained models such as BERT or GPT, leading to faster inference and lower deployment costs. Compared to fine-tuning or distillation, pruning has the added advantage of not requiring external knowledge transfer, making it well-suited for real-time or low-resource scenarios such as mobile applications and online question-answering systems. Overall, pruning compresses the model structure while preserving accuracy, effectively balancing the trade-off between model size and performance.

\subsubsection{Quantization}

Quantization is a widely adopted model compression and acceleration technique that converts high-precision floating-point representations into lower-precision integer formats (The third	 column in Figure \ref{fig:lightweight}). The key idea is that high-precision (e.g., 32-bit floating point) representations of model parameters are not always necessary—particularly during inference—where reduced-precision representations can significantly decrease computational burden and memory consumption while maintaining acceptable performance. By quantizing a model, its weights and activations can be represented using 8-bit integers or even lower-precision formats, thereby enabling efficient computation and fast response in resource-constrained environments.

Quantization techniques are typically classified into three main categories: static quantization, dynamic quantization, and mixed-precision quantization. Static quantization applies quantization to the model parameters before deployment, and is suitable for scenarios where consistent, low-precision inference is required. Dynamic quantization, on the other hand, quantizes activations at runtime, offering greater flexibility and reduced accuracy degradation, particularly for applications with variable or unpredictable input distributions. Mixed-precision quantization assigns different levels of numerical precision to different parts of the model—for example, using low precision for weights while maintaining higher precision for certain activation layers—to achieve a more optimal trade-off between accuracy and efficiency.

By converting models from high-precision floating-point formats to low-precision integers, quantization significantly reduces computational overhead, making it especially effective for deployment on mobile devices, embedded systems, and edge computing platforms. Quantization accelerates inference speed, lowers power consumption, and reduces memory usage—all without altering the underlying model structure. Compared with techniques such as distillation or pruning, quantization offers higher efficiency and lower overhead, making it particularly suitable for real-time inference, on-device deployment, and low-latency applications.

In summary, quantization provides a cost-effective and flexible optimization path for deploying large-scale pretrained models. It strikes a practical balance between inference efficiency and resource utilization, thereby extending the applicability of foundation models to a broader range of real-world geophysical scenarios.

\subsubsection{Application of RAG in Foundation Model Deployment}

Retrieval-Augmented Generation (RAG) is a knowledge-enhanced architecture that integrates information retrieval and text generation to improve performance on knowledge-intensive tasks. Its core objective is to supplement large language models with external knowledge sources, thereby enhancing their accuracy, factual consistency, and adaptability to dynamic information environments. The fundamental architecture of RAG consists of two main components: a retriever and a generator. When given a query or prompt, the retriever searches a large external knowledge base—such as Wikipedia, domain-specific corpora, or structured databases—for semantically relevant documents or passages. The retrieval process can rely on either dense retrieval or sparse retrieval techniques. In dense retrieval, both queries and documents are encoded into high-dimensional vector representations, and similarity (typically cosine similarity) is computed in the embedding space to identify the most relevant candidates. In contrast, sparse retrieval methods (e.g., TF-IDF or BM25) rely on keyword frequency statistics for relevance matching.

The retrieved documents are then passed, together with the original query, to the generator, which is typically a pretrained language model (e.g., GPT, T5). The generator synthesizes the final output by incorporating the retrieved context, thereby grounding the response in external evidence. This mechanism allows RAG to combine the generalization ability of generative models with up-to-date and task-specific knowledge from external sources, effectively overcoming limitations in model size, domain coverage, and temporal validity. Furthermore, RAG supports end-to-end training, enabling the retriever and generator to be optimized jointly for task-specific performance. This cooperative training enhances the alignment between the retrieval and generation components, leading to improved coherence and relevance in the output. The RAG architecture has demonstrated strong performance in a range of applications, including question answering, dialogue generation, and knowledge reasoning, offering significant gains in accuracy, factual consistency, and knowledge richness. As such, it serves as a powerful paradigm for deploying large-scale models in real-world knowledge-driven scenarios.

The workflow of RAG (Retrieval-Augmented Generation) is an end-to-end knowledge-enhanced generation process, centered on the tight integration of the retrieval and generation modules to dynamically incorporate external knowledge and enhance performance on knowledge-intensive tasks. The process begins with an input in the form of a question or prompt, which is first encoded by the retriever module into either a dense vector or a sparse representation. The retriever then queries a large-scale external knowledge base—such as Wikipedia, domain-specific documents, or structured data sources—to retrieve semantically relevant documents or passages.

The retrieval step typically employs either dense retrieval or sparse retrieval techniques. In dense retrieval, the input and all candidate documents are embedded into a high-dimensional vector space, and cosine similarity is used to select the most relevant candidates. In sparse retrieval, matching is performed using term-frequency-based statistical methods, such as TF-IDF or BM25, which rank documents based on lexical overlap and relevance scores.

The retrieved documents are then passed, along with the original input, to the generator module. The generator, typically a pretrained language model (e.g., GPT, T5), generates the final answer or text output based on both the input query and the contextual information retrieved. The generator is capable not only of understanding the semantics of the input question, but also of integrating retrieved external knowledge for reasoning and response generation, thereby producing more accurate and content-rich outputs. This entire pipeline is optimized via an end-to-end training framework, enabling joint learning between the retriever and generator. Such coordination ensures that the retriever selects content that is most useful for the generator, thereby enhancing the model’s overall effectiveness in tasks such as question answering, dialogue generation, and knowledge-based reasoning.

By fully leveraging the language understanding and generation capabilities of large models, while dynamically retrieving external information, the RAG framework effectively addresses limitations in model knowledge coverage and temporal sensitivity. This leads to significant improvements in the accuracy, relevance, and knowledge richness of the generated responses.

\subsubsection{Application of Agent Technology in Foundation Model Deployment}

Agent technology, originating from the field of artificial intelligence (AI), refers to a class of software entities capable of autonomously perceiving their environment, learning, making decisions, and executing tasks. Agents are characterized by several key properties: autonomy, intelligence, interactivity, and adaptability. Autonomy allows an agent to accomplish assigned tasks without human intervention. Intelligence enables the agent to make context-aware decisions based on changes in its environment. Interactivity and adaptability empower the agent to learn continuously and refine its behavior to respond effectively to complex and dynamic scenarios.

These capabilities make agent-based systems highly suitable for deployment in environments that require continuous decision-making, real-time adaptation, and task-specific optimization—all of which are critical for unlocking the full potential of large-scale foundation models in real-world applications.

The integration of geophysical foundation models with agent technology holds significant promise for advancing modern geophysical exploration, particularly in enhancing capabilities in data processing and decision support. Geophysical foundation models—especially those based on deep learning and large-scale pretraining—possess powerful abilities in feature extraction and pattern recognition, offering considerable advantages in handling complex geophysical data.

By embedding these models within agent-based systems, the intelligence and autonomy of various exploration tasks—such as task allocation, data analysis, anomaly detection, and strategic decision-making—can be greatly improved.

In practical terms, geophysical foundation models can be incorporated directly into the decision-making pipeline of agents, enabling them to make faster and more informed decisions when confronted with new exploration objectives. For example, an agent equipped with deep learning-based seismic analysis capabilities can autonomously process seismic data, extract potential subsurface structural features, and formulate adaptive exploration strategies based on those features.

Moreover, agents often need to handle heterogeneous geophysical data types, including seismic, gravity, and magnetic field measurements. Foundation models, through multi-task learning, can facilitate knowledge sharing across these tasks, improving overall model generalization. Agents can dynamically select and adapt appropriate models based on the input data type and task requirements, thus increasing both the precision and efficiency of the exploration process.

The integration of reinforcement learning with geophysical foundation models further endows agents with the ability for adaptive learning and real-time decision-making throughout the exploration process. For example, when exploration outcomes deviate from expectations, the agent—through its connection with the foundation model—can dynamically update model parameters to achieve self-adaptive optimization. In this way, the agent not only learns from data but also makes flexible decisions in previously unseen or uncertain environments.

Geophysical foundation models also play a crucial role in data fusion, particularly in the integration of multi-source geophysical datasets. Agents can construct geophysics-based knowledge graphs to systematically fuse heterogeneous data types. These knowledge graphs, when combined with foundation models, enable the agent to perform advanced tasks such as anomaly detection and trend prediction through graph reasoning and semantic inference.

As both geophysics and artificial intelligence continue to advance, the synergistic integration of foundation models and agent technology is expected to play an increasingly pivotal role across multiple dimensions. In the future, foundation model-enabled agents will drive innovations in automated exploration and intelligent decision-making, allowing the entire workflow—from data acquisition and analysis to predictive modeling—to be completed with minimal human intervention.

Through continued refinement of deep learning and reinforcement learning techniques, agents will be capable of offering highly accurate and efficient decision support in complex exploration scenarios. Moreover, as geophysical exploration inherently spans multiple scientific domains, the integration of foundation models into agent architectures will promote cross-disciplinary knowledge sharing and collaboration. By embedding domain-adaptive foundation models into intelligent agents, exploration tasks can be carried out in a coordinated and scalable manner across diverse fields, accelerating the advancement of next-generation geophysical technologies.

With the advancement of real-time data processing technologies, agents empowered by foundation models will enable faster responses and more accurate predictions during geophysical exploration. This will allow exploration strategies to be dynamically adjusted based on real-time monitoring and forecasting, significantly enhancing operational efficiency. In summary, the integration of geophysical foundation models with agent technology is poised to play a pivotal role in advancing the intelligent automation of exploration workflows. As these technologies continue to mature, AI-driven intelligent agents will become indispensable tools in geophysical exploration—improving precision, efficiency, and the degree of automation across a wide range of applications.

\section{Discussion}\label{sec:capabilities}

\subsection{Future perspectives of foundation model}

The future development of foundation models in geophysics holds tremendous potential to revolutionize how data is collected, analyzed, and applied across the discipline. As foundation model technology continues to evolve, it is expected to drive transformative changes in numerous subfields of geophysics—particularly in data integration, resource exploration, disaster early warning, geological modeling, automated data processing, and the realization of digital earth platforms.

\begin{figure*}[!htb]%
\centering
\includegraphics[width=1\textwidth]{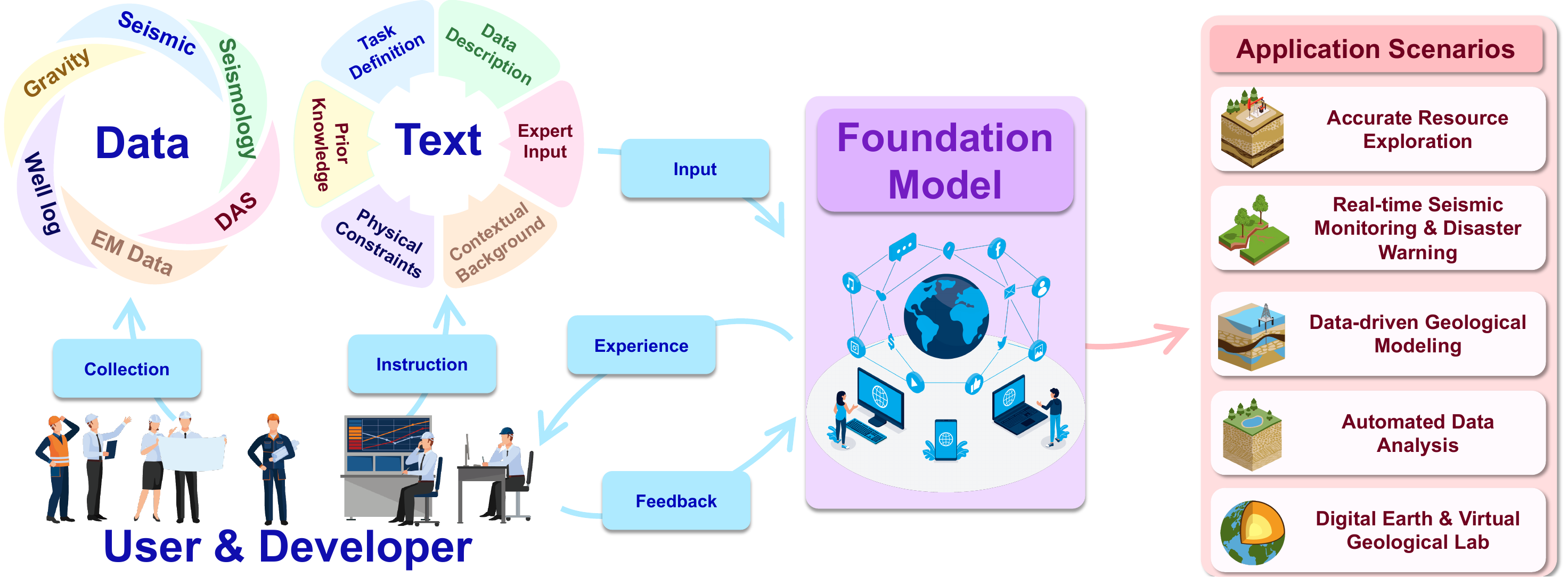}
\caption{Final outcomes and future vision of geophysical foundation models. This figure illustrates how foundation models can integrate multimodal data—such as seismic, gravity, electromagnetic, and well logging data—with textual information including task definitions, data descriptions, expert inputs, and physical constraints, to drive forward the development of geophysical science. These models hold promise across a wide range of application scenarios, including precise resource exploration, real-time seismic monitoring and disaster early warning, data-driven geological modeling, automated data analysis, and the creation of digital earth platforms and virtual geological laboratories.
}
\label{fig:final}
\end{figure*}

\subsubsection{Multidimensional Inputs at the Inference Stage}

\paragraph{Textual Inputs}
Textual input plays a crucial role in foundation models. It can be used not only for task definition (e.g., specifying the geophysical problem to be addressed, such as exploration objectives or disaster warning indicators) but also for data description (e.g., data sources, acquisition methods, formats), expert feedback (e.g., domain-driven corrections or experiential knowledge), physical constraints (e.g., fundamental geophysical laws such as the wave equation, gravitational and magnetic field equations), prior knowledge (e.g., previous exploration results or geological interpretations), and contextual information (e.g., task background, operational scenarios).

By leveraging textual input, foundation models are able to better understand the task context and objectives, allowing for flexible adaptation across diverse geophysical applications. Text-based interfaces also enable iterative optimization of model outputs, as experts can inject knowledge, refine model behavior, and guide inference through natural language interactions. The model, in turn, can adjust dynamically in response to new knowledge and feedback.

\paragraph{Multidisciplinary Data}
Geophysical investigations often require the integration of multi-source heterogeneous datasets, including seismic, gravity, magnetic, and electrical methods. These data vary significantly in terms of format, structure, and physical meaning, necessitating unified interfaces and preprocessing protocols to enable seamless fusion within a single modeling framework.

Foundation models must be designed to accommodate the unique characteristics of each data modality, including sampling strategies, spatial-temporal resolution, and measurement uncertainty. Proper handling of these factors is essential to ensure efficient and accurate data fusion, which in turn supports robust, cross-disciplinary geophysical interpretation. Ultimately, enabling multidisciplinary data to coexist and interact within foundation models will significantly expand the scope and precision of geophysical analysis.

\subsubsection{Core Mechanisms of Foundation Models}

\paragraph{Mixture of Experts (MoE) Routing Capability}
Foundation models can leverage the Mixture of Experts (MoE) mechanism to dynamically allocate computational resources and modeling capacity based on the characteristics of specific tasks and datasets. In geophysical applications, where task diversity is significant, the model can automatically select the most appropriate expert sub-model for each task. For example, seismic data inversion can be routed to a specialized expert trained on wave propagation, while gravitational anomaly detection may be handled by a gravity-focused expert module. This task-aware expert routing enables not only efficient specialization but also joint optimization across experts, thereby enhancing overall model performance across a range of geophysical tasks.

\paragraph{User Feedback and Adaptive Optimization}
Geophysical modeling often requires iterative feedback and correction from domain experts. Foundation models can incorporate interactive user feedback mechanisms to adjust predictions in real time. Feedback can be provided through natural language instructions or other interactive modalities, enabling flexible task refinement. Examples include:

\begin{itemize}
\item \textbf{Inversion parameter adjustment}: Experts can fine-tune model parameters such as regularization terms or loss functions based on field data or laboratory results. \item \textbf{Objective redefinition}: If exploration objectives shift, experts can redefine the task's objective function, prompting the model to recalibrate its optimization target. \item \textbf{Multi-round iterative refinement}: Throughout the task lifecycle, experts can provide continuous feedback, enabling the model to iteratively self-optimize and progressively improve output accuracy. 
\end{itemize}

In addition, foundation models can incorporate techniques such as incremental learning and transfer learning to continuously adapt based on new data and feedback. This allows for ongoing refinement of model parameters and inference outcomes, ensuring that the model remains flexible and performant across evolving geophysical scenarios and user requirements.

\subsubsection{Future Outlook}

\paragraph{Resource Exploration}
Foundation models will play a transformative role in helping geophysicists more accurately identify subsurface resources such as hydrocarbons and mineral deposits, thereby improving the efficiency and precision of exploration efforts. By fusing data from diverse sources—including seismic, gravity, magnetic, and electrical surveys—foundation models can rapidly generate spatial predictions of potential resource distributions along with probabilistic assessments of their likelihood. This capability reduces the complexity of manual interpretation and lowers the overall cost of exploration.

As model capabilities continue to advance, foundation models will be able to autonomously discover novel exploration targets from raw or minimally processed data, offering valuable insights that support informed decision-making. Ultimately, such advancements will contribute to the sustainable development and responsible utilization of energy and mineral resources.

\paragraph{Disaster Early Warning}

Foundation models will also play a critical role in natural disaster early warning systems, particularly in the context of high-impact events such as earthquakes and volcanic eruptions. By integrating real-time monitoring data—including seismic waveforms, ground deformation measurements, and satellite observations—foundation models can perform immediate analysis to provide early warnings and forecast disaster trends.

These capabilities will enhance the detection of potential hazard signals and deliver actionable intelligence to emergency response agencies, enabling timely and effective interventions. In doing so, foundation models will help minimize the human and economic losses associated with natural disasters and improve societal resilience in vulnerable regions.

\paragraph{Geological Modeling}

With the continuous growth of data availability and improvements in model precision, foundation models are expected to play an increasingly pivotal role in geological modeling. These models can automatically or semi-automatically integrate a wide range of geophysical data sources—such as seismic, gravity, magnetic, and electrical measurements—to generate more accurate and dynamic 3D or 4D geological models.

Such models will enable geoscientists to gain deeper insights into subsurface structures, tectonic features, and reservoir distributions, thereby providing robust support for hydrocarbon and mineral exploration. Additionally, foundation models can be customized for different application scenarios, adapting to varying geological settings and exploration objectives.

\paragraph{Automated Data Processing}

The application of foundation models in data processing will significantly streamline geophysical workflows and improve operational efficiency. In traditional geophysical studies, tasks such as data cleaning, format conversion, and anomaly detection are often labor-intensive and time-consuming.

Foundation models can automate these processes—from raw data preprocessing and feature extraction to visualization of analytical results—thereby reducing manual workload and increasing overall throughput. This not only saves considerable labor costs but also enhances the accuracy, consistency, and reliability of geophysical data analysis.

\paragraph{Digital Earth}

Looking ahead, foundation models are expected to serve as a critical technological pillar for the construction of Digital Earth platforms. By integrating global-scale geophysical data, remote sensing imagery, and geographic information systems (GIS) into a unified framework, foundation models enable dynamic simulation and real-time monitoring of Earth systems.

Digital Earth not only provides a novel paradigm for scientific research but also holds tremendous potential in domains such as resource management, environmental monitoring, and urban planning. Foundation models will offer robust data support and decision-making assistance for the sustainable development of Digital Earth, empowering society to better understand and respond to global environmental changes.

\paragraph{Conclusion}

As foundation models continue to evolve within the geophysical sciences, their applications in resource exploration, disaster early warning, geological modeling, automated data processing, and Digital Earth development are expected to expand rapidly. These advancements will significantly enhance both the research and practical dimensions of geophysics, contributing to societal progress and environmental sustainability.

The future of foundation models is rich with promise and has the potential to bring transformative change to Earth science, reshaping the way we observe, analyze, and interact with the planet.

\subsection{Open-Source Culture and Collaboration}

In the rapid advancement of deep learning and large-scale models, open-source culture has played a pivotal role. Open-source initiatives not only foster technological sharing and innovation, but also accelerate research progress across numerous disciplines—including geophysics. The evolution of deep learning in geophysical applications has greatly benefited from open-source frameworks such as TensorFlow and PyTorch, which provide researchers with powerful tools and extensive algorithmic support. These frameworks facilitate the efficient processing of large-scale data and offer a wide range of predefined model architectures and optimization strategies, which are particularly valuable for developing deep learning algorithms tailored to geophysical data.

In geophysics, datasets are often extremely large and characterized by high complexity and heterogeneity, making data analysis inherently challenging. Open-source platforms enable researchers to leverage pre-built models and algorithms, which can be directly applied to tasks such as seismic waveform inversion, gravity and magnetic data interpretation, and more. Moreover, the sharing of open-source code allows researchers to rapidly adapt, extend, and optimize successful methods developed in other domains. This significantly shortens the development cycle and empowers researchers to achieve meaningful results within a shorter time frame.

The sharing of pretrained models is another critical contribution of open-source culture. In the deep learning community, many research teams openly release their pretrained models, significantly accelerating the application, replication, and refinement of advanced techniques. This practice is equally valuable in geophysics. By sharing and fine-tuning pretrained models, researchers can rapidly experiment with and analyze domain-specific geophysical datasets.

For instance, pretrained Convolutional Neural Networks (CNNs) or Recurrent Neural Networks (RNNs) can be adapted for processing seismic data, thereby expediting tasks such as automatic seismic wave identification and inversion. Model sharing not only lowers research and development costs, but also facilitates widespread adoption of deep learning methods, ultimately advancing innovation and collaboration in the geophysical sciences.

\subsection{Hardware Computation and Cost Analysis}

From GPT-3 to the latest GPT-5 and LLaMA 3 series, the hardware requirements for training large-scale models have increased dramatically~\cite{Thompson2024}. For instance, GPT-3, completed in 2020, was trained using 10,000 NVIDIA V100 GPUs, requiring a total of 3,552,000 GPU hours at an estimated cost of \$9 million USD.

In contrast, GPT-5, trained in 2024, utilized 50,000 NVIDIA H100 GPUs, with a total training time of 144,000,000 GPU hours, and an estimated cost reaching \$612 million USD. This sharp increase in computational cost reflects not only the advancement of GPU performance, but also the exponential growth in model scale and complexity.

\begin{table}[!htbp]
\centering
\begin{tabular}{|l|l|l|l|}
\hline
\textbf{Chip Type} & \textbf{\makecell{Pricing \\Date}} & \textbf{\makecell{\$ per \\Chip-Hour}} & \textbf{1M Hours (\$)} \\ 
\hline
V100  & 2020 & \$0.66 & \$660,000 \\ 
\hline
V100  & 2020 & \$2.48 & \$2,480,000 \\ 
\hline
A100  & 2023 & \$3.93 & \$3,930,000 \\ 
\hline
H100  & 2023 & \$4.25 & \$4,250,000 \\ 
\hline
TPUv4 & 2023 & \$3.22 & \$3,220,000 \\ 
\hline
TPUv5e & 2024 & \$1.20 & \$1,200,000 \\ 
\hline
TPUv5p & 2024 & \$4.20 & \$4,200,000 \\ 
\hline
\end{tabular}
\caption{GPU/TPU Chip Pricing and Cost for 1M Hours~\cite{Thompson2024}}
\label{tab:gpu_pricing}
\end{table}

Moreover, there are significant differences in cost-effectiveness across hardware platforms. For example, TPUs (Tensor Processing Units) and GPUs offer different trade-offs in terms of computational density and efficiency. In the case of Gemini, trained in 2023, TPUv4 demonstrated exceptionally high performance and cost-efficiency. The Gemini model was trained using 57,000 TPUv4 units, totaling 136,800,000 TPU hours, at an estimated cost of \$440 million USD.

In contrast, the widespread adoption of NVIDIA H100 GPUs in 2024 has provided a versatile alternative for high-performance model training. For instance, the LLaMA 3 model with 405 billion parameters was trained using 24,576 H100 GPUs, with a total cost of approximately \$125 million USD~\cite{Thompson2024}. These examples highlight the evolving hardware landscape and the critical role of infrastructure selection in managing training cost and efficiency for large-scale models.

\begin{table*}[!htbp]
\centering
\begin{tabular}{|p{1.5cm}|p{1.5cm}|p{1cm}|p{1.5cm}|p{1cm}|p{1cm}|p{1.5cm}|p{1.5cm}|p{1.5cm}|}
\hline
\textbf{Model} & \textbf{Training End} & \textbf{Chip \newline Type} & \textbf{TFLOP/s \newline (max)} & \textbf{Chip \newline Count} & \textbf{Wall Clock Time \newline (days)} & \textbf{Total Time \newline (hours)} & \textbf{Total Time \newline (years)} & \textbf{Retail Cost \newline (\$US)} \\ 
\hline
GPT-3          & Apr/2020             & V100               & 130                    & 10,000              & 15                           & 3,552,000                   & 405                         & \$9M                        \\ 
\hline
Llama 1        & Jan/2023             & A100               & 312                    & 2,048               & 21                           & 1,032,192                   & 118                         & \$4M                        \\ 
\hline
Llama 2        & Jun/2023             & A100               & 312                    & 2,048               & 35                           & 1,720,320                   & 196                         & \$7M                        \\ 
\hline
Titan          & Apr/2023             & A100               & 312                    & 13,760              & 48                           & 11,558,400                  & 1,319                       & \$45M                       \\ 
\hline
GPT-4          & Aug/2022             & A100               & 312                    & 25,000              & 95                           & 57,000,000                  & 6,503                       & \$224M                      \\ 
\hline
Gemini         & Nov/2023             & TPUv4             & 275                    & 57,000              & 100                          & 136,800,000                 & 15,606                      & \$440M                      \\ 
\hline
Llama 3 70B    & Apr/2024             & H100               & 989                    & 24,576              & 11                           & 6,300,000                   & 719                         & \$7M                        \\ 
\hline
Llama 3 405B   & Apr/2024             & H100               & 989                    & 24,576              & 50                           & 29,491,200                  & 3,364                       & \$125M                      \\ 
\hline
GPT-5          & Mar/2024             & H100               & 989                    & 50,000              & 120                          & 144,000,000                 & 16,428                      & \$612M                      \\ 
\hline
Olympus        & Aug/2024             & H100               & 989                    & -                   & -                            & -                           & -                           & -                           \\ 
\hline
Grok 2         & Jun/2024             & H100               & 989                    & 20,000              & 50                           & 57,600,000                  & 6,571                       & \$245M                      \\ 
\hline
Gemini 2       & Nov/2024             & TPUv6             & 1847                   & -                   & -                            & -                           & -                           & -                           \\ 
\hline
Grok 3         & Dec/2024             & H100               & 989                    & 100,000             & 50                           & 288,000,000                 & 32,855                      & \$1.2B                      \\ 
\hline
\end{tabular}
\caption{Training Configurations and Costs for Large Models~\cite{Thompson2024}}
\label{tab:training_costs}
\end{table*}

\subsection{Distributed Computing and Architecture Optimization}

As model sizes expand from billions to tens of trillions of parameters, distributed computing has become an essential strategy for scalable training. Techniques such as model parallelism, data parallelism, and pipeline parallelism enable the efficient allocation of computational resources and help alleviate hardware bottlenecks. For example, the training of GPT-4 utilized 25,000 NVIDIA A100 GPUs and relied on distributed algorithms to complete 95 days of intensive training. This form of architecture optimization significantly reduced training time and improved hardware utilization.

Next-generation models—such as Gemini 2 and Grok 3 (both expected in 2024)—are characterized by even more advanced distributed architectures and will likely depend on TPUv6 and H100 GPUs to meet their computational demands. The training of Grok 3 is projected to require 100,000 H100 GPUs, consuming 288,000,000 GPU hours, with a total estimated cost of \$1.2 billion USD.

Such extraordinary compute requirements highlight the growing importance of distributed systems complexity and optimization strategies as a critical area of future research. Efficient parallelization schemes, memory management, fault tolerance, and interconnect bandwidth will play increasingly vital roles in enabling the next frontier of large-scale AI and scientific modeling.

\subsection{Research Funding}

With the rise of large-scale pretrained models, both academia and industry face growing demands for substantial research and development investments. Constructing foundation models for geophysical applications requires not only extensive data acquisition and processing, but also significant computational resources and technical infrastructure. These projects are often large in scale, encompassing all stages—from data collection and annotation to model training and optimization. Thus, financial support is indispensable for advancing this field.

Geophysical data typically originate from multiple complex sources, including seismic waveforms, gravity measurements, and electromagnetic field data. Acquiring and processing such data is often cost-intensive, particularly in challenging environments such as deep-sea or remote regions, where specialized equipment and logistical support are required. In addition, training deep learning models at scale in geophysics demands massive computing resources, often involving high-performance GPU or TPU clusters.

Consequently, funding from national agencies, industry stakeholders, and academic institutions is crucial to the success of these initiatives. Such investment ensures the efficient acquisition and processing of geophysical data and provides the infrastructure and technical support necessary for the development of high-capacity models. Continued funding is essential to bridge the gap between traditional geophysical practices and cutting-edge AI-driven exploration.

\subsection{Data Privacy and Security}

With the widespread adoption of big data and artificial intelligence in the geophysical sciences, issues related to ethics, privacy, and data governance have become increasingly prominent. Geophysical data often contain extensive environmental and subsurface information, and their collection, processing, and use must adhere to strict ethical standards and privacy protection protocols. This concern is particularly critical when the data involve sensitive geographic regions or personally identifiable information, making compliance and security essential challenges to address.

Data privacy is a non-negligible issue in the geophysical domain. Many datasets, especially those related to resource exploration or environmental monitoring, may contain confidential or sensitive information. To ensure legality and compliance, researchers are expected to implement safeguards such as data encryption and de-identification, thereby protecting sensitive content from unauthorized access or misuse. Furthermore, data users are obligated to follow principles of transparency, fairness, and accountability throughout data collection, sharing, and usage, to safeguard the rights and interests of all stakeholders involved.

On the other hand, as large-scale pretrained models become increasingly integrated into geophysical applications, ethical concerns surrounding the models themselves are also emerging. For instance, how can biases in seismic data processing be mitigated? How can the model's fairness and reliability be ensured across diverse geological settings? To address these challenges, it is essential to embed ethical review mechanisms throughout the model development lifecycle. This includes ensuring diversity and representativeness in dataset selection and annotation, and validating model outputs to prevent bias introduced by data imbalance or overfitting.

By implementing these measures, researchers can promote fairness, transparency, and responsible AI deployment in the geophysical sciences, ultimately ensuring that foundation models contribute to scientifically sound and ethically grounded outcomes.

\hypertarget{conclusion}{\section{Conclusion}}
\label{sec:conclusion}
This paper presents a comprehensive review of the development and application of foundation models in geophysics, offering a systematic framework that encompasses the full pipeline—from geophysical data acquisition to model deployment. We thoroughly analyzed key components including data collection and preprocessing, model architecture design, pretraining strategies, and deployment methodologies. In particular, we proposed tailored solutions to address the inherent diversity, complexity, and physical consistency requirements of geophysical data.

Our study fills a critical gap in the current literature by providing the first holistic overview of foundation model development in the geophysical domain. It also advances the practical application of foundation models in geophysical research, especially by improving computational efficiency, reducing dependence on labeled data, and enhancing model interpretability.

Throughout the review, we identified and discussed key challenges in applying foundation models to geophysical problems, such as how to effectively handle unlabeled data, mitigate physical inconsistency, and address the computational demands of large-scale training. We further explored how transfer learning and self-supervised learning strategies inherent to foundation models can offer more efficient solutions for domain-specific tasks.

Looking forward, future research should focus on expanding the application of foundation models to more complex and interdisciplinary geophysical problems, with particular attention to multi-modal data integration, physics-informed learning, and explainable AI. As data volumes continue to grow and computational capabilities advance, foundation models are poised to drive innovative breakthroughs in geophysics.

We believe that the continued development of foundation models will play an increasingly critical role in geophysical research, providing powerful technical support for more precise subsurface imaging, resource exploration, and environmental protection.

\bibliography{referList}

\bibliographystyle{ACM-Reference-Format}



\end{document}